\documentclass[11pt]{article}

\usepackage{SIGACTNews}

\newcommand{\COLUMNsplit}{\vskip 2em\hrule \vskip 3em}

\newcommand{\COLUMNguest}[2]{%
  \begin{center}%
   {\LARGE\bf #1 \par}%
    \vskip 1.5em%
    {\Large
      \lineskip .75em%
      \begin{tabular}[t]{c}%
        #2
      \end{tabular}\par}%
    \vskip 1.5em%
  \end{center}%
  \par}

\usepackage{epsfig}
\usepackage{amssymb}
\usepackage{times}
\usepackage{comment}
\usepackage{amsthm}

\newtheorem{theorem}	 			{Theorem}[section]
\newtheorem*{theoremnonumber}                   {Theorem}
\newtheorem{lemma}		[theorem]	{Lemma}

\newtheorem{definition}		[theorem]	{Definition} 
\newtheorem{prop}		[theorem]	{Proposition}

\newtheorem{exercise}  [theorem]  {Exercise}
\newtheorem{examplecore}[theorem]{Example}

\newenvironment{example}
  {\begin{examplecore}\rm}
  {\hfill $\Box$\end{examplecore}}

\newenvironment{pf}{\noindent\textbf{Proof\/}.}{\hfill $\Box$ \vspace{1mm}}

\newcommand{\tup}[1]{\overline{#1}}

\newcommand{\qcsp}{\mathsf{QCSP}}

\newcommand{\csp}{\mathsf{CSP}}

\newcommand{\pol}{\mathsf{Pol}}
\newcommand{\spol}{\mathsf{sPol}}
\newcommand{\inv}{\mathsf{Inv}}

\newcommand{\la}{\langle}
\newcommand{\ra}{\rangle}

\newcommand{\cons}{\ensuremath{\mathcal{C}}}

\newcommand{\majority}{\ensuremath{\mathsf{majority}}}
\newcommand{\minority}{\ensuremath{\mathsf{minority}}}

\newcommand{\false}{\mathsf{false}}
\newcommand{\true}{\mathsf{true}}

\newcommand{\prefix}{\mathcal{P}}

\title{SIGACT News Logic Column 17}
\author{Riccardo Pucella\\
Northeastern University\\
Boston, MA 02115 USA\\
riccardo@ccs.neu.edu}
\date{}

\begin{document}

\SIGACTmaketitle

I am always looking for contributions. If you have any suggestion
concerning the content of the Logic Column, or if you would like to
contribute by writing a column yourself, feel free to get in touch
with me.

\COLUMNsplit

\COLUMNguest{A Rendezvous of Logic, Complexity, and Algebra\footnote{\copyright{} Hubie Chen, 2006.}}
   {Hubie Chen\\Universitat Pompeu Fabra}

\section{Pop quiz}
\label{section:popquiz}

Recall the \emph{propositional satisfiability} ({\sc SAT}) problem: 
we are given a propositional formula such as
$$(s \vee t) \wedge (\neg s) \wedge (\neg u \vee s \vee \neg t) \wedge (\neg s \vee t)$$
consisting of a conjunction of \emph{clauses}, where a clause is a disjunction
of literals; a literal is either a variable $v$ 
(a \emph{positive} literal)
or the negation of a variable
$\neg v$ (a \emph{negative} literal).  
We are to decide if there is an assignment to the variables satisfying
the formula, that is, an assignment under which every clause contains
at least one true literal.
The example formula is satisfied by the assignment $f$ where
$f(s) = f(u) = \false$ and $f(t) = \true$.
The {\sc SAT} problem is famously regarded as the first natural problem
to be identified as NP-complete.

Two special cases of the {\sc SAT} problem that are well-known to be
decidable in polynomial time are the \emph{{\sc 2-SAT}} problem, 
in which every clause is a \emph{2-clause}, 
a clause having exactly two literals, 
as in the formula
$$(\neg u \vee v) \wedge (\neg u \vee \neg v) \wedge (\neg v \vee w) \wedge (\neg w \vee t) \wedge (\neg t \vee v)$$
and the \emph{{\sc Horn-SAT}} problem, 
in which every clause is a \emph{Horn clause},
a clause having at most one positive literal,
as in the formula
$$(\neg y \vee x_1) \wedge (\neg y' \vee \neg x_1 \vee y) \wedge (\neg x_2 \vee \neg y) \wedge (\neg x_1 \vee x_2).$$

Now consider the {\sc Quantified SAT} problem.  
We are given a quantified
formula such as 
$$\forall v \forall t \exists u \exists w ((\neg u \vee v) \wedge (\neg u \vee \neg v) \wedge (\neg v \vee w) \wedge (\neg w \vee t) \wedge (\neg t \vee v))$$
or
$$\forall y \forall y' \forall y'' \exists x_1 \exists x_2 ((\neg y \vee x_1) \wedge (\neg y' \vee \neg x_1 \vee y) \wedge (\neg x_2 \vee \neg y) \wedge (\neg y'' \vee \neg x_1 \vee x_2)),$$
that is, a formula consisting of 
a conjunction of clauses preceded by a quantifier prefix
in which all of the variables are quantified.
Our task is to decide if the formula is true or false.

In general, {\sc Quantified SAT} is known to be PSPACE-complete.
However, we can very well consider the special cases 
{\sc Quantified 2-SAT} and {\sc Quantified Horn-SAT},
where the clauses are restricted to be 2-clauses and Horn clauses,
respectively.
These two special cases of {\sc Quantified SAT} are known to be
polynomial-time decidable~\cite{apt79,kbs87}.  
Let us focus on $\Pi_2$ (or ``$\forall \exists$'')
formulas,
that is, quantified formulas where the quantifier prefix
consists of a sequence of universally quantified variables followed
by a sequence of existentially quantified variables 
(as in the above two examples).
I claim that there is a \emph{simple} proof that the $\Pi_2$ formulas
of {\sc Quantified 2-SAT} and {\sc Quantified Horn-SAT} 
are polynomial-time tractable.  
More precisely, I claim that
there are simple reductions from {\sc Quantified 2-SAT} to {\sc 2-SAT}
and from {\sc Quantified Horn-SAT} to {\sc Horn-SAT} that can be 
justified by short proofs.
At this point, I would like to kindly ask the reader to stop reading,
and attempt to demonstrate this.

\paragraph{\bf Warning: spoilers ahead.}
I now present the claimed reductions and accompanying proofs.
Let us begin with {\sc Quantified Horn-SAT}.
What do we want to do?  We are given a $\Pi_2$ formula
$$\Phi = \forall y_1 \ldots \forall y_m \exists x_1 \ldots \exists x_n \phi$$
where $\phi$ is the conjunction of Horn clauses, and we want to 
efficiently decide if $\Phi$ is true.  
Let $Y_{\Phi}$ denote the universally quantified variables of $\Phi$,
and let $X_{\Phi}$ denote the existentially quantified variables of $\Phi$.
Here, because the formula $\Phi$ has prefix
class $\Pi_2$, we may observe that it is true if and only if
for every assignment
$f: Y_{\Phi} \rightarrow \{ \true, \false \}$,
there exists an
extension $f': Y_{\Phi} \cup X_{\Phi} \rightarrow \{ \true, \false \}$ 
of $f$ under which the clauses $\phi$ is true.
How might we check this property?  

Given a \emph{single} assignment $f: Y_{\Phi} \rightarrow \{ \true, \false \}$,
we can certainly check efficiently whether or not it has an extension
$f': Y_{\Phi} \cup X_{\Phi} \rightarrow \{ \true, \false \}$ 
under which $\phi$ is true: we simply instantiate the universally
quantified variables $Y_{\Phi}$ according to $f$, and then use
any polynomial-time algorithm for {\sc Horn-SAT} to decide if the resulting
{\sc Horn-SAT} formula is satisfiable.
However, if we are to act in polynomial time,
we definitely do not have time to perform this check for
\emph{all} assignments to the universally quantified variables, as there
are $2^{|Y_{\Phi}|}$ such assignments---too many!

Interestingly enough, it turns out that it suffices to perform the
``extension check'' for a restricted set of assignments, in order
to determine truth of the formula $\Phi$.
For an integer $j \geq 1$ and a constant $b \in \{ \true, \false \}$,
let $[\leq j, b]_{\Phi}$ denote the set of all assignments
$f: Y_{\Phi} \rightarrow \{ \true, \false \}$ to the universally
quantified variables of $\Phi$ such that the number of variables
mapped to $b$ is less than or equal to $j$.
For example, $[\leq 1, \false]_{\Phi}$ contains the assignment
sending all variables in $Y_{\Phi}$ to $\true$, and all assignments
on $Y_{\Phi}$ in which exactly one variable is sent to $\false$.
I claim that $\Phi$ is true as long as
 all assignments in $[\leq 1, \false]_{\Phi}$
have a satisfying extension.

\begin{prop}
\label{prop:gen-qhornsat}
Let $\Phi$ be an instance of {\sc Quantified Horn-SAT} having prefix class 
$\Pi_2$.
The formula $\Phi$ is true if and only if for every assignment
$f \in [\leq 1, \false]_{\Phi}$, there exists an extension
$f': Y_{\Phi} \cup X_{\Phi} \rightarrow \{ \true, \false \}$ of $f$
satisfying all clauses of $\Phi$.
\end{prop}

Clearly, this proposition yields the correctness of the following
procedure for deciding a formula $\Phi$ from our class: 
for every assignment $f \in [\leq 1, \false]_{\Phi}$, 
instantiate the variables $Y_{\Phi}$ according to $f$ and use a polynomial-time
algorithm for {\sc Horn-SAT} to check if the resulting clauses are satisfiable;
if they are satisfiable for every assignment $f$, return ``true'', otherwise,
return ``false''.
Since---relative to the size of $\Phi$---there 
are polynomially many (in fact, linearly many) assignments
in $[\leq 1, \false]_{\Phi}$ this procedure
is indeed a polynomial-time procedure.

In order to prove the proposition, we will make use of the following concept.
Say that a propositional formula $\psi$ is \emph{preserved} by an operation
$g: \{ \true, \false \}^k \rightarrow \{ \true, \false \}$
if for any $k$ assignments $f_1, \ldots, f_k$ under which $\psi$
is true, the formula $\psi$ is also true under
the assignment $g(f_1, \ldots, f_k)$ defined by
$(g(f_1, \ldots, f_k))(v) = g(f_1(v), \ldots, f_k(v))$ for all variables $v$.

\begin{example}
\label{ex:and-preserves-any-horn-clause}
We can verify
 that the boolean AND operation $\wedge$ preserves any Horn clause.
Let $l_1 \vee \ldots \vee l_c$ be a Horn clause, where the $l_i$ denote
literals, and let $v_i$ denote the variable underlying the literal $l_i$.
Let $f_1, f_2$ be assignments under which the clause is true.
If one of the assignments $f_1, f_2$ satisfies a negative literal
$l_i = \neg v_i$, then $(\wedge(f_1, f_2))(v_i) = \false$
and the clause is true under $\wedge(f_1, f_2)$.
Otherwise, every negative literal is false under both $f_1$ and $f_2$,
and since $f_1, f_2$ are satisfying assignments, 
there must be a positive literal $l_j = v_j$ with 
$f_1(v_j) = f_2(v_j) = \true$.  Then $(\wedge(f_1, f_2))(v_j) = \true$
and the clause is true under $\wedge(f_1, f_2)$.
\end{example}

\begin{example}
\label{ex:conjunction}
Now let $\phi = C_1 \wedge \ldots \wedge C_l$ be a conjunction of
clauses for which there is an operation 
$g: \{ \true, \false \}^k \rightarrow \{ \true, \false \}$
preserving each clause $C_i$.  We can verify that $\phi$ itself is preserved
by $g$.  Indeed, let $f_1, \ldots, f_k$ be any assignments
under which $\phi$ is true.  Then, consider any clause $C_i$.
The clause $C_i$ is true under all of the assignments
$f_1, \ldots, f_k$.  Since $g$ preserves $C_i$ we have that $C_i$
is true under $g(f_1, \ldots, f_k)$, 
and since our choice of $C_i$ was arbitrary
we have that $\phi$ is true under $g(f_1, \ldots, f_k)$.
\end{example}

From these two examples, we see that every conjunction of Horn clauses
is preserved by the operation~$\wedge$.
With this fact in hand, we may now turn to the proof of the proposition.

\begin{pf} (Proposition~\ref{prop:gen-qhornsat})
The ``only if'' direction is clear, so we prove the ``if'' direction.
Let $m$ denote $|Y_{\Phi}|$.
We prove by induction that the following holds for all $i = 1, \ldots, m$:
every assignment $f \in [\leq i, \false]_{\Phi}$ has an extension
$f': Y_{\Phi} \cup X_{\Phi} \rightarrow \{ \true, \false \}$
satisfying the clauses $\phi$ of $\Phi$.
This suffices, since $[\leq m, \false]_{\Phi}$ is the set of all assignments
to $Y_{\Phi}$.

The base case $i = 1$ holds by hypothesis.

Suppose that $i \geq 2$.
Let $f$ be an assignment in $[\leq i, \false]_{\Phi}$.
If $f \in [\leq i-1, \false]_{\Phi}$ then the desired extension $f'$
exists by induction.
So suppose that $f$ maps exactly $i$ variables $S = \{ s_1, \ldots, s_i \}$
to the value $\false$.
Let $f_1: Y_{\Phi} \rightarrow \{ \true, \false \}$ 
be the assignment mapping exactly the variables 
$S \setminus \{ s_1 \}$ to $\false$, and
let $f_2: Y_{\Phi} \rightarrow \{ \true, \false \}$ 
be the assignment mapping exactly the variables 
$S \setminus \{ s_2 \}$ to $\false$.
Since $f_1, f_2 \in [\leq i-1, \false]_{\Phi}$,
they have extensions $f'_1, f'_2$ satisfying $\phi$.
Since $\wedge$ preserves $\phi$, the assignment $f' = \wedge(f'_1, f'_2)$
also satisfies $\phi$.  
The assignment $f'$ is an extension of $f$: 
\begin{itemize}

\item $f'(s_1) = \wedge(f'_1(s_1), \false) = \false$

\item $f'(s_2) = \wedge(\false, f'_2(s_2)) = \false$

\item for all $s \in S \setminus \{ s_1, s_2 \}$, 
$f'(s) = \wedge(\false, \false) = \false$

\item for all $y \in Y_{\Phi} \setminus S$, 
$f'(y) = \wedge(\true, \true) = \true$

\end{itemize}
\end{pf}

Looking now at {\sc Quantified 2-SAT}, we give a proof of tractability
(again, for $\Pi_2$ formulas) that is similar in spirit to the proof
we just gave for {\sc Quantified Horn-SAT}.
Whereas Horn clauses were preserved by the operation $\wedge$,
2-clauses are preserved by a different operation.
Let $\majority$ denote the ternary operation
defined by
$\majority(x, y, z) = (x \wedge y) \vee (x \wedge z) \vee (y \wedge z)$.
That is, $\majority$ returns the input value occurring
at least twice.
It is straightforward to verify
 that $\majority$ preserves any 
2-clause.\footnote{
Here is a verification:
 let $l_1 \vee l_2$ be a 2-clause, 
let $f_1, f_2, f_3$ be assignments satisfying this clause, 
and let $v_1, v_2$
denote the variables underlying the literals $l_1, l_2$, respectively.
If two of the assignments $f_1, f_2, f_3$ are equal on $\{ v_1, v_2 \}$, 
then $\majority(f_1, f_2, f_3)$ is equal to those two assignments
on $\{ v_1, v_2 \}$ and hence satisfies the clause.
Otherwise, $f_1$, $f_2$, and $f_3$, restricted to $\{ v_1, v_2 \}$,
are exactly the three distinct assignments satisfying $l_1 \vee l_2$,
and it is can be seen that both literals $l_1, l_2$
are true under $\majority(f_1, f_2, f_3)$.
}
Having observed this, we can now establish a result similar in spirit to
Proposition~\ref{prop:gen-qhornsat}.

\begin{prop}
\label{prop:gen-q2sat}
Let $\Phi$ be 
an instance of {\sc Quantified 2-SAT} having prefix class 
$\Pi_2$.
The formula $\Phi$ is true if and only if for every assignment
$f \in [\leq 2, \false]_{\Phi}$, there exists an extension
$f': Y_{\Phi} \cup X_{\Phi} \rightarrow \{ \true, \false \}$ of $f$
satisfying all clauses of $\Phi$.
\end{prop}

As with Proposition~\ref{prop:gen-qhornsat}, we can readily infer
a polynomial-time algorithm for the quantified formulas under study
from Proposition~\ref{prop:gen-q2sat}.
Namely, using a polynomial-time algorithm for {\sc 2-SAT},
it can be decided whether or not
 for all $f \in [\leq 2, \false]_{\Phi}$, the desired
extension $f'$ exists.  Since there are polynomially many assignments
in $[\leq 2, \false]_{\Phi}$, this can be carried out in polynomial time.

\begin{pf}
This proof is structurally identical to the proof of
Proposition~\ref{prop:gen-qhornsat}.
As in that proof, 
the ``only if'' direction is clear, so we prove the ``if'' direction.
Let $m$ denote $|Y_{\Phi}|$.
We prove by induction that the following holds for all $i = 2, \ldots, m$:
every assignment $f \in [\leq i, \false]_{\Phi}$ has an extension
$f': Y_{\Phi} \cup X_{\Phi} \rightarrow \{ \true, \false \}$
satisfying the clauses $\phi$ of $\Phi$.
This suffices, since $[\leq m, \false]_{\Phi}$ is the set of all assignments
to $Y_{\Phi}$.

The base case $i = 2$ holds by hypothesis.

Suppose that $i \geq 3$.
Let $f$ be an assignment in $[\leq i, \false]_{\Phi}$.
If $f \in [\leq i-1, \false]_{\Phi}$ then the desired extension $f'$
exists by induction.
So suppose that $f$ maps exactly $i$ variables $S = \{ s_1, \ldots, s_i \}$
to the value $\false$.
For $j \in \{ 1, 2, 3 \}$, let $f_j: Y_{\Phi} \rightarrow \{ \true, \false \}$ 
be the assignment
mapping exactly the variables 
$S \setminus \{ s_j \}$ to $\false$.
Since the assignments $f_1, f_2, f_3$ are in $[\leq i-1, \false]_{\Phi}$,
they have extensions $f'_1, f'_2, f'_3$ satisfying $\phi$.
Since $\majority$ preserves any 2-clause, by
the discussion in Example~\ref{ex:conjunction} it also preserves
$\phi$, 
and thus the assignment $f' = \majority(f'_1, f'_2, f'_3)$
also satisfies $\phi$.  
The assignment $f'$ is an extension of $f$: 
\begin{itemize}

\item $f'(s_1) = \majority(f'_1(s_1), \false, \false) = \false$

\item $f'(s_2) = \majority(\false, f'_2(s_2), \false) = \false$

\item $f'(s_3) = \majority(\false, \false, f'_3(s_3)) = \false$

\item for all $s \in S \setminus \{ s_1, s_2, s_3 \}$, 
$f'(s) = \majority(\false, \false, \false) = \false$

\item for all $y \in Y_{\Phi} \setminus S$, 
$f'(y) = \majority(\true, \true, \true) = \true$

\end{itemize}
\end{pf}

I mention that the polynomial-time tractability of the
special cases of {\sc Quantified SAT} that have been discussed,
\emph{without} the $\Pi_2$ restriction,
 is proved using the notion of preservation by an operation
 in~\cite{chen06-collapsibility}.

\section{What's this all about?}

What just happened?  The question of how to efficiently decide the truth
of certain logical formulas was posed, and answered by considering operations
preserving the formulas.
This situation exemplifies a theme underlying an emerging line of research
that studies the complexity of \emph{constraint satisfaction problems}.

\paragraph{\bf More details on that, please.} 
The constraint satisfaction problem (CSP) 
is a general framework in which
many search problems can be readily modeled; 
an instance of the CSP consists of a set of constraints on variables,
and the goal is to determine if there is an assignment to the variables
satisfying every one of the given constraints.
A broad family of problems 
can be obtained from the CSP 
by restricting the so-called \emph{constraint language}---a set of
relations that can be used to form constraints.
Each constraint language $\Gamma$
gives rise to a particular computational problem, denoted by
$\csp(\Gamma)$, and a focal research question is to describe the
complexity of $\csp(\Gamma)$ for all constraint languages $\Gamma$.
The family of problems $\csp(\Gamma)$ is extremely rich, and includes
graph homomorphism problems, the problem of solving a system
of equations over various algebraic structures, and the problems
{\sc SAT}, {\sc 2-SAT} and {\sc Horn-SAT}.
The research area studying the problems $\csp(\Gamma)$ has recognized that
a set of operations---an \emph{algebra}---can be associated to each
constraint language $\Gamma$
in such a way that information on the complexity of
the problem $\csp(\Gamma)$
can be derived from these operations.  This association has given way to 
a fruitful interaction among the areas of logic, complexity, and algebra.

\paragraph{\bf What of the results presented in the previous section?}
Both Propositions~\ref{prop:gen-qhornsat} 
and \ref{prop:gen-q2sat}
have been known.  For example, they were proved by
Gr\"{a}del~\cite{gradel92} to obtain results in descriptive complexity;
Proposition~\ref{prop:gen-qhornsat} was also derived
 by Karpinski et al.~\cite{kbs87} from theory establishing the tractability
of {\sc Quantified Horn-SAT}.
In those papers these propositions were derived by arguments
strongly based on the clausal forms of the formulas under study, in contrast
to the algebraic arguments given here based on \emph{operations}.
I hope the reader will agree that the proofs given here are
particularly short and simple, and yield evidence that the utilized algebraic
viewpoint can shed light even on classically studied fragments
of propositional logic.  
I believe it is also worth emphasizing
that the proofs of Propositions~\ref{prop:gen-qhornsat} 
and~\ref{prop:gen-q2sat} are structurally identical,
and thus that the reasoning employed in both cases
is generic and not heavily tied to the particular formulas under study.

\paragraph{What happens in this article?}
In this expository article, I give a contemporary, algebraic treatment
of the inaugural result on constraint languages,
Schaefer's theorem on 
boolean constraint satisfaction problems~\cite{schaefer78}.
This theorem classifies the complexity of $\csp(\Gamma)$
for all constraint languages $\Gamma$ over a \emph{two}-element domain.
In particular, it gives a description of the constraint languages
$\Gamma$ such that $\csp(\Gamma)$ is polynomial-time tractable,
and shows that for all other constraint languages $\Gamma$,
the problem $\csp(\Gamma)$ is NP-complete.
This theorem is of broad interest, as it provides a rich
class of NP-complete boolean satisfiability problems,
some of which have extremely simple descriptions,
and which---as Schaefer himself envisioned---often facilitate
the development of a NP-hardness proof.
Following an introduction to the algebraic viewpoint
on constraint satisfaction (Section~\ref{section:polymorphisms}),
I give a relatively short but complete proof of Schaefer's theorem
(Sections~\ref{section:tractability} and~\ref{section:intractability}).
After this, I prove---again using algebraic techniques---an 
analog of Schaefer's theorem
for quantified satisfiability problems
(Section~\ref{section:quantified}),
and then give a ``fine'' classification theorem
for quantified satisfiability problems
where the number of quantifier alternations is bounded
 (Section~\ref{section:bd-alternation}).
In the last section of this paper (Section~\ref{section:to-infinity}),
I discuss a recently discovered application of the algebraic viewpoint
to a class of logical formulas falling outside the framework of
constraint satisfaction, and open the question of finding more results
of this form.

Almost all of the results presented here appear either explicitly
or implicitly in the literature: the exposition draws 
upon many previous publications, including
the works of Post~\cite{post41} and Rosenberg~\cite{rosenberg83-fivetypes}
on clone theory;
the papers of Geiger~\cite{geiger68} and Bodnarchuk et al.~\cite{bodnarchuk69} 
identifying
a relevant Galois connection;
the paper of Schaefer~\cite{schaefer78}; 
the papers of
Jeavons, Cohen, and Gyssens~\cite{jcg97}, Jeavons~\cite{jeavons98},
and Bulatov, Jeavons, and Krokhin~\cite{bjk05} connecting constraint
satisfaction to clone theory and universal algebra;
and the tractability result of Jeavons, Cohen, and Cooper~\cite{jcc98}.
The novelty here is that
I attempt to give a unified and relatively short account of these results,
and include a number of lesser-known proofs.

Care has been taken to make the presentation of this article
self-contained, and I do not assume any background in any of the areas
touched by this article, other than familiarity with basic
complexity-theoretic notions such as 
reducibility and the complexity classes P and NP.
I hope that a wide variety of readers will be able to
take something away from this article---from those who would
like to understand (a nice version of) the statement of Schaefer's theorem
and catch a glimpse of its inner workings, to
those interested in incorporating the presented ideas and techniques into
their research toolboxes.
Indeed, I have endeavored to give a streamlined proof of Schaefer's
theorem and the other classification theorems, and believe
that the presented proof of Schaefer's theorem could reasonably
be taught in a few lectures of a course, 
under the assumption of a mature audience.

Throughout this article, I provide exercises for the reader;
some are relatively routine, while others offer a taste of deeper ideas.
To close this section, the following exercises expanding upon
the discussion in Section~\ref{section:popquiz} are offered.

\begin{exercise}
Define the operation $\minority$ to be the the ternary operation
such that $\minority(x, y, z) = x \oplus y \oplus z$, where
$\oplus$ denotes the usual exclusive OR operation.  That is, if
all of the inputs to $\minority$ are the same value, that value is the output;
otherwise, $\minority$ returns whichever one of its inputs occurs exactly once.
Show that any equation of the form 
$v_1 \oplus \cdots \oplus v_k = c$ where the $v_i$ are variables
and $c \in \{ 0, 1 \}$ is a constant, is preserved by $\minority$.
\end{exercise}

\begin{exercise}
Using the previous exercise, prove a result analogous to
Propositions~\ref{prop:gen-qhornsat} and~\ref{prop:gen-q2sat}
for $\Pi_2$ formulas where the
quantifier-free part is the conjunction of equations
of the form described in the previous exercise.
\end{exercise}

\begin{exercise}
Show that Proposition~\ref{prop:gen-q2sat} holds with 
$([\leq 1, \false]_{\Phi} \cup [\leq 0, \true]_{\Phi})$
in place of
$[\leq 2, \false]_{\Phi}$.
Observe that the set of assignments 
$([\leq 1, \false]_{\Phi} \cup [\leq 0, \true]_{\Phi})$
is in general smaller than
$[\leq 2, \false]_{\Phi}$!
Also observe that, by duality,
this proposition holds with
$([\leq 1, \true]_{\Phi} \cup [\leq 0, \false]_{\Phi})$ or
$[\leq 2, \true]_{\Phi}$ in place of
$[\leq 2, \false]_{\Phi}$.

\end{exercise}

\section{Constraint satisfaction and polymorphisms}
\label{section:polymorphisms}

In this section, we describe the computational problems of interest 
and the algebraic tools that will be used to study them.

\begin{definition}
A \emph{relation} over domain $D$ is a subset of $D^k$ for some $k \geq 1$;
$k$ is said to be the \emph{arity} of the relation.
A \emph{constraint language} over domain $D$ is a set of relations
over $D$.
A constraint language is \emph{finite} if it contains finitely many relations,
and is \emph{boolean} if it is over the two-element domain $\{ 0, 1 \}$.
\end{definition}

By a \emph{domain} $D$, we simply mean a set.
Other than in the last section, we 
will be concerned primarily with
 constraint languages over a finite domain $D$.
Also, note that we will use $0$ and $1$ to denote the boolean values
$\false$ and $\true$.

\begin{definition}
A \emph{constraint} over a constraint language
$\Gamma$ is an expression of the form
$R(v_1, \ldots, v_k)$ where $R$ is a relation of arity $k$ 
contained in $\Gamma$,
and the $v_i$ are variables.  A constraint is satisfied by a mapping
$f$ defined on the $v_i$ if $(f(v_1), \ldots, f(v_k)) \in R$.
\end{definition}

The computational problems we are interested in are defined as follows.

\begin{definition}
Let $\Gamma$ be a finite constraint language over domain $D$.
The problem $\csp(\Gamma)$ is to decide, given a finite set of variables $V$
and a finite set of constraints over $\Gamma$ with variables from $V$,
whether or not there exists a 
\emph{solution} (or \emph{satisfying assignment}), a
mapping $f: V \rightarrow D$
satisfying all of the constraints.
\end{definition}

Observe that, for all constraint languages $\Gamma$, the problem
$\csp(\Gamma)$ is in NP: a variable assignment $f: V \rightarrow D$
has polynomial size, and whether or not it satisfies all constraints
can be checked in polynomial time.
We remark that when discussing problems of the form $\csp(\Gamma)$,
we will confine our attention to \emph{finite} constraint languages $\Gamma$.
This permits us to avoid certain technicalities and
discussion of how relations are represented, although we should note that
the complexity of 
infinite constraint languages is considered in the literature.

\begin{example}
\label{ex:3-sat}
We demonstrate that {\sc 3-SAT}, the case of the {\sc SAT} problem
where every clause has exactly three literals, can be viewed as a problem
of the form $\csp(\Gamma)$ for a boolean constraint language $\Gamma$.  
Define the relations
$R_{0,3}$, $R_{1,3}$, $R_{2,3}$, and $R_{3,3}$ by
\begin{center}
$
\begin{array}{cccccccccccccc}
R_{0,3} & = & \{ 0, 1 \}^3  \setminus  \{ (0, 0, 0) \}\\
R_{1,3} & = & \{ 0, 1 \}^3  \setminus  \{ (1, 0, 0) \}\\ 
R_{2,3} & = & \{ 0, 1 \}^3  \setminus  \{ (1, 1, 0) \}\\ 
R_{3,3} & = & \{ 0, 1 \}^3  \setminus  \{ (1, 1, 1) \}\\
\end{array}
$
\end{center}
Notice that for any variables $x, y, z$, we have the following equivalences:
\begin{center}
$
\begin{array}{lllllllll}
R_{0,3}(x,y,z) & \equiv & (x \vee y \vee z)\\
R_{1,3}(x,y,z) & \equiv & (\neg x \vee y \vee z)\\
R_{2,3}(x,y,z) & \equiv & (\neg x \vee \neg y \vee z)\\
R_{3,3}(x,y,z) & \equiv & (\neg x \vee \neg y \vee \neg z)\\
\end{array}
$
\end{center}
That is, (as an example)
the constraint $R_{1,3}(x,y,z)$ is satisfied by an assignment
if and only if the clause $(\neg x \vee y \vee z)$ is satisfied by
the assignment.

Let $\Gamma_3$ be the constraint language 
$\{ R_{0,3}, R_{1,3}, R_{2,3}, R_{3,3} \}$.
Every instance of the {\sc 3-SAT} problem can be readily translated
into an instance of $\csp(\Gamma_3)$ having the same satisfying assignments.  
For instance, consider the {\sc 3-SAT} instance
$$(\neg u \vee s \vee \neg t) \wedge (\neg s \vee t \vee v) \wedge
(s \vee t \vee \neg v) \wedge (v \vee u \vee s).$$
It is equivalent to the $\csp(\Gamma_3)$ instance
with variables $\{ s, t, u, v \}$ and constraints
$$\{ R_{2,3}(u, t, s), R_{1,3}(s, t, v), R_{1,3}(v, s, t), R_{0,3}(v, u, s) \}.$$
Similarly, any instance of $\csp(\Gamma_3)$ can be formulated as an instance
of {\sc 3-SAT}.
\end{example}

It is well-known that 3-SAT is NP-hard; we formulate this as follows.

\begin{prop}
\label{prop:3-sat-np-hard}
The problem $\csp(\Gamma_3)$, where $\Gamma_3$ as is defined in 
Example~\ref{ex:3-sat}, is NP-hard.
\end{prop}

\newcommand{\nae}{R_{\mathsf{NAE}}}

\begin{example}
\label{ex:nae}
Schaefer~\cite{schaefer78} identified the problem
{\sc Not-all-equal satisfiability}: given a collection of
sets $S_1, \ldots, S_m$
each having at most $3$ members, can the members be colored with
two colors so that no set is all one color?
We show that this problem is equivalent to $\csp(\Gamma)$ where
$\Gamma$ contains the single relation 
$\nae = \{ 0, 1 \}^3 \setminus \{ (0, 0, 0), (1, 1, 1) \}$.

Take an instance $S_1, \ldots, S_m$ of {\sc Not-all-equal satisfiability};
we translate it to an instance of $\csp(\Gamma)$ by creating, for each
set $S_i$, a constraint $\nae(s, t, u)$ where $s, t, u$ are such that
$S_i = \{ s, t, u \}$.  It is readily seen that a coloring
$f: (\cup_{i=1}^m S_i) \rightarrow \{ 0, 1 \}$ satisfies the condition
given in the problem description if and only if it satisfies
all of the constraints.  Similarly, an instance of $\csp(\Gamma)$
can be translated to an instance of {\sc Not-all-equal satisfiability}
by creating, for each constraint $\nae(s, t, u)$, a set 
$\{ s, t, u \}$.
\end{example}

We now give a notion of definability for relations.  As we will see
momentarily, this notion will permit a constraint language to
``simulate'' relations that might not be inside the constraint language.

\begin{definition}
We say that a relation $R \subseteq D^k$ is \emph{pp-definable}
(short for \emph{primitive positive definable}) from 
a constraint language $\Gamma$ if for some $m \geq 0$
there exists a finite conjunction $\cons$ consisting of
constraints and  equalities $(u = v)$ 
over variables $\{ v_1, \ldots, v_k, x_1, \ldots, x_m \}$
 such that
$$R(v_1, \ldots, v_k) \equiv \exists x_1 \ldots \exists x_m \cons.$$
That is, $R$ contains exactly those tuples of the form
$(g(v_1), \ldots, g(v_k))$ where $g$ is an assignment that can be extended
to a satisfying assignment of $\cons$.
We use $\la \Gamma \ra$ to denote the set of all relations that are
pp-definable from $\Gamma$.
\end{definition}

\begin{example}
Let $S = \{ (0, 1), (1, 0) \}$ be the disequality relation over
$\{ 0, 1 \}$.  The following is a pp-definition of $S$ from
the constraint language $\Gamma_3$ of 
Example~\ref{ex:3-sat}:
$$S(y,z) = \exists x (R_{0,3}(x, y, z) \wedge R_{1,3}(x, y, z) \wedge R_{2, 3}(z, y, x) \wedge R_{3, 3}(z, y, x)).$$
\end{example}

When all relations in a constraint language $\Gamma'$ are 
pp-definable in another constraint language $\Gamma$, we have that
the constraint satisfaction problem 
over $\Gamma'$ reduces to that over $\Gamma$.

\begin{prop} (implicit in~\cite{jeavons98})
\label{prop:expressivity-reduction}
Let $\Gamma$ and $\Gamma'$ be finite constraint languages.
If $\Gamma' \subseteq \la \Gamma \ra$, then 
$\csp(\Gamma')$ reduces to $\csp(\Gamma)$.
\end{prop}

Note that the only notion of reduction used in this article
is many-one polynomial-time reduction.\footnote{
We remark that, by making use of the result of Reingold~\cite{Reingold},
the reduction of Proposition~\ref{prop:expressivity-reduction}
can be carried out in logarithmic space.
}

\begin{pf}
From an instance $\phi$ of $\csp(\Gamma')$, we create an instance of
$\csp(\Gamma)$ in the following way.  
We loop over each constraint
$C = R(v_1, \ldots, v_k)$ in
$\phi$, performing the following operations for each:
let $\exists x_1 \ldots \exists x_m \cons$ be a pp-definition of $C$
over $\Gamma$, rename the existentially quantified variables
$x_1, \ldots, x_m$ if necessary so that they are distinct from
all variables of other constraints, and replace $C$ with 
the constraints in $\cons$.
It is clear that each replacement preserves the satisfiability of the
CSP instance; moreover, since there are finitely many relations in 
$\Gamma'$, each has a pp-definition of constant size, and 
all of the replacements can be carried out in polynomial time.
The result is a set of constraints over $\Gamma$ and
equalities; each equality $(u = v)$ can be eliminated by 
removing it from the set and 
replacing all instances of (say) $v$ with $u$.
\end{pf}

From this proposition, it can be seen that a finite constraint language
$\Gamma$ is tractable if and only if all finite subsets $\Gamma'$
of $\la \Gamma \ra$ are tractable: the forward direction follows
from the proposition, while the backwards direction follows by taking
$\Gamma' = \Gamma$.
This observation can be interpreted as saying that the tractability
of a constraint language $\Gamma$ is characterized by the set 
$\la \Gamma \ra$, and justifies focusing on the sets $\la \Gamma \ra$.
Interestingly, we will show that the set of relations
$\la \Gamma \ra$ is in turn characterized by a set of operations called
the \emph{polymorphisms} of $\Gamma$.

\begin{definition}
\label{def:polymorphism}
An operation $f: D^m \rightarrow D$ is a \emph{polymorphism} of
a relation $R \subseteq D^k$ if for any choice of $m$ tuples
$(t_{11}, \ldots, t_{1k}), \ldots, (t_{m1}, \ldots, t_{mk})$ from $R$,
it holds that the tuple obtained from these $m$ tuples by applying
$f$ coordinate-wise,
$(f(t_{11}, \ldots, t_{m1}), \ldots, f(t_{1k}, \ldots, t_{mk}))$, is
in $R$.
\end{definition}

That is, an operation $f$ is a polymorphism of a relation $R$ if 
$R$ satisfies a closure property: applying $f$ to any tuples in $R$
yields another tuple inside $R$.
This notion of closure is essentially equivalent to that
required by the notion of preservation used in Section~\ref{section:popquiz};
however, whereas there we spoke of \emph{formulas} being preserved by
operations, here we speak of \emph{relations} having polymorphisms.
We now give some examples, which consider
the constraint language
$\Gamma_3 = \{ R_{0,3}, R_{1,3}, R_{2,3}, R_{3,3} \}$ from
Example~\ref{ex:3-sat}.

\begin{example}
The relation $R_{0,3}$ has the boolean OR $\vee$ operation as a polymorphism;
we can see this as follows.
Suppose that
 $(t_{11}, t_{12}, t_{13}), (t_{21}, t_{22}, t_{23})$ are two tuples
from $R_{0,3}$.
There is some coordinate of the first tuple equal to $1$, that is,
there exists $j \in \{ 1, 2, 3 \}$
such that $t_{1j} = 1$.
It follows that $t_{1j} \vee t_{2j} = 1$, and thus the tuple
$(t_{11} \vee t_{21}, t_{12} \vee t_{22}, t_{13} \vee t_{23})$
is also contained in $R_{0,3}$.

The relation $R_{1,3}$ also has the boolean OR $\vee$ operation as a
polymorphism.  
Let us take two tuples
 $(t_{11}, t_{12}, t_{13}), (t_{21}, t_{22}, t_{23})$ 
from $R_{1,3}$.  If the first tuple $(t_{11}, t_{12}, t_{13})$
is equal to $(0, 0, 0)$,
the OR of the two tuples
$(t_{11} \vee t_{21}, t_{12} \vee t_{22}, t_{13} \vee t_{23})$
is equal to the second tuple, which is contained in $R$ by assumption.
Otherwise, either $t_{12}$ or $t_{13}$ is equal to $1$, implying that
one of the values
$(t_{12} \vee t_{22})$,  $(t_{13} \vee t_{23})$ is equal to $1$ and that
$(t_{11} \vee t_{21}, t_{12} \vee t_{22}, t_{13} \vee t_{23})$
is contained in $R_{1,3}$.

The relation $R_{2,3}$ does not have the boolean OR $\vee$ operation
as a polymorphism.
This is because the two tuples $(1, 0, 0)$, $(0, 1, 0)$ are both in
$R_{2,3}$, but their OR, the tuple $(1, 1, 0)$, is not.
\end{example}

\begin{example}
None of the relations in 
$\Gamma_3 = \{ R_{0,3}, R_{1,3}, R_{2,3}, R_{3,3} \}$
have the $\majority$ operation as a polymorphism.
Indeed, let $R$ be any relation of the form
$\{ 0, 1 \}^3 \setminus \{ (a_1, a_2, a_3) \}$
with $a_1, a_2, a_3 \in \{ 0, 1 \}$.
Observe that
the tuples
$(\neg a_1, a_2, a_3)$
$(a_1, \neg a_2, a_3)$
$(a_1, a_2, \neg a_3)$
are all contained in $R$, but applying the $\majority$ operation to
these tuples yields the tuple
$(\majority(\neg a_1, a_1, a_1), 
  \majority(a_2, \neg a_2, a_2), 
  \majority(a_3, a_3, \neg a_3))$
which is equal to $(a_1, a_2, a_3)$ and is hence not in $R$.
\end{example}

Upon initial acquaintance,
 the notion of polymorphism may
appear unrelated to the notion of pp-definability.
Actually, it turns out that
the polymorphisms of a constraint language $\Gamma$ contain
enough information to derive the set of relations $\la \Gamma \ra$!
We say that an operation $f: D^m \rightarrow D$ is a \emph{polymorphism}
of a constraint language $\Gamma$ if it is a polymorphism of all relations
$R \in \Gamma$, and we use $\pol(\Gamma)$ to denote the set of all
polymorphisms of $\Gamma$, that is,
$$\pol(\Gamma) = \{ f: \forall R \in \Gamma, f \mbox{ is a polymorphism of } R \}.$$
  Also, for a set of operations $O$,
we use $\inv(O)$ to denote the set of relations having all operations
in $O$ as a polymorphism, that is, 
$$\inv(O) = \{ R: \forall f \in O, f \mbox{ is a polymorphism of } R \}.$$

\begin{theorem}
\label{thm:galois-connection}
Let $\Gamma$ be a finite constraint language over a finite domain $D$.
It holds that $\la \Gamma \ra = \inv( \pol( \Gamma ))$.
\end{theorem}

In words, this theorem states that
 a relation is pp-definable from $\Gamma$ exactly when all 
polymorphisms of $\Gamma$ are polymorphisms of it.
Again, this result shows that the set of relations $\la \Gamma \ra$
can be derived from the set of operations $\pol(\Gamma)$, in particular,
by applying the $\inv(\cdot)$ operator.
This theorem was established by
Geiger and Bodnarchuk et al.~\cite{geiger68,bodnarchuk69}.\footnote{
We remark that the $\pol(\cdot)$ and $\inv(\cdot)$ operators
give rise to an instance of a \emph{Galois connection}.
}
The proof of the $\supseteq$ direction given here
is based on 
a proof that appeared in Dalmau's Ph.D. thesis~\cite{dalmau00-thesis}.

\begin{pf}
We first show that $\la \Gamma \ra \subseteq \inv( \pol( \Gamma))$;
this is the more straightforward direction.
Let 
$$R(v_1, \ldots, v_k) \equiv \exists x_1 \ldots \exists x_m \cons$$
be the pp-definition of a relation $R$ over $\Gamma$.
Suppose that $g: D^n \rightarrow D$ is a polymorphism of $\Gamma$;
we want to show that $g$ is a polymorphism of $R$.

Consider first the relation $R'$ defined by 
$$R'(v_1, \ldots, v_k, x_1, \ldots, x_m) \equiv \cons.$$
Let $\tup{t_1}, \ldots, \tup{t_n}$ be tuples in $R'$.
Each tuple $\tup{t_i}$ has the form
$(f_i(v_1), \ldots, f_i(v_k), f_i(x_1), \ldots, f_i(x_m))$
for an assignment $f_i$ satisfying $\cons$.
Let $S(w_1, \ldots, w_l)$ be any constraint or equality of $\cons$.
We have that $(f_i(w_1), \ldots, f_i(w_l)) \in S$ for each $f_i$.
Since $S$ has $g$ as a polymorphism, we have that the arity $l$ tuple
with 
$g(f_1(w_i), \ldots, f_n(w_i))$ as its $i$th coordinate
is in $S$.  Thus the mapping sending each variable 
$v \in \{ v_1, \ldots, v_k, x_1, \ldots, x_m \}$ to 
$g(f_1(v), \ldots, f_n(v))$, satisfies all constraints of $\cons$.
We then have that the
the tuple $g(\tup{t_1}, \ldots, \tup{t_n})$, where $g$ is applied
coordinate-wise, is in $R'$, and thus that $g$ is a polymorphism of
$R'$.
(This is essentially the argument of Example~\ref{ex:conjunction},
but in slightly different language.)

Now, we have that $R'$ has $g$ as a polymorphism, and want to show that
$$R(v_1, \ldots, v_k) \equiv \exists x_1 \ldots \exists x_m R'(v_1, \ldots, v_k, x_1, \ldots, x_m)$$
has $g$ as a polymorphism.
Let $\tup{t_1}, \ldots, \tup{t_n}$ be tuples in $R$.  
Each tuple 
$\tup{t_i} = (t_{i1}, \ldots, t_{ik})$ 
has an extension
$\tup{t'_i} = (t_{i1}, \ldots, t_{i(k+m)})$ contained in $R'$.
We want to show that the tuple
$g(\tup{t_1}, \ldots, \tup{t_n})$, where $g$ is applied coordinate-wise,
has an extension in $R'$.  The tuple
$g(\tup{t'_1}, \ldots, \tup{t'_n})$ is such an extension.

We now prove $\la \Gamma \ra \supseteq \inv( \pol( \Gamma ))$.
Suppose that $R \in \inv( \pol( \Gamma ))$.
Let $n$ denote the arity of $R$, 
let $m$ denote the number of tuples in $R$, 
and let
$(t_{11}, \ldots, t_{1n}), \ldots, (t_{m1}, \ldots, t_{mn})$ denote
the tuples of $R$ (in any order).
We may assume that there are no distinct coordinates $i, j$ such that
$(t_{1i}, \ldots, t_{mi}) = (t_{1j}, \ldots, t_{mj})$, as
these may be eliminated one by one using the following observation:
let $\sigma(1), \ldots, \sigma(n-1)$ denote the sequence $1, \ldots, n$
with $j$ removed; then, if
$\exists x_1 \ldots \exists x_m \cons$ is a pp-definition for the
relation $R'$ defined by
$R'(v_{\sigma(1)}, \ldots, v_{\sigma(n-1)}) \equiv
\exists v_j (R(v_1, \ldots, v_n))$,
we have that $\exists x_1 \ldots \exists x_m (\cons \wedge (v_i = v_j))$
is a pp-definition for $R(v_1, \ldots, v_n)$.

We create a conjunction of constraints $\cons$ over the variable set $D^m$.
Our conjunction $\cons$ contains, for each relation $S \in \Gamma$
and sequence of $m$ tuples
$(s_{11}, \ldots, s_{1k}), \ldots, (s_{m1}, \ldots, s_{mk}) \in S$,
a constraint
$S((s_{11}, \ldots, s_{m1}), \ldots, (s_{1k}, \ldots, s_{mk}))$;
here, $k$ denotes the arity of $S$.
It is straightforward to verify that an assignment
$f: D^m \rightarrow D$ satisfies all of the constraints in $\cons$
if and only if it is an $m$-ary polymorphism of $\Gamma$.

Now consider the relation $R'$ defined by
$$R'((t_{11}, \ldots, t_{m1}), \ldots, (t_{1n}, \ldots, t_{mn})) = 
\exists v_1 \ldots \exists v_p \cons$$
where $v_1, \ldots, v_p$ are the tuples in 
$D^m \setminus 
\{ (t_{11}, \ldots, t_{m1}), \ldots, (t_{1n}, \ldots, t_{mn}) \}$,
in any order.
Since it is exactly the $m$-ary polymorphisms of $\Gamma$ that satisfy
$\cons$, we have
$$R' = \{ (f(t_{11}, \ldots, t_{m1}), \ldots, f(t_{1n}, \ldots, t_{mn})): f \in \pol(\Gamma), f \mbox{ has arity } m \}.$$
We claim that $R' = R$, which yields the proof.  

$R' \subseteq R$: The tuples
$(t_{11}, \ldots, t_{1n}), \ldots, (t_{m1}, \ldots, t_{mn})$ are
in $R$; since every $f \in \pol(\Gamma)$ is a polymorphism of $R$, 
it follows that 
$
(f(t_{11}, \ldots, t_{m1}), \ldots, f(t_{1n}, \ldots, t_{mn})) \in R$.

$R \subseteq R'$: Let $\pi_i: D^m \rightarrow D$ denote the function
that projects onto the $i$th coordinate, that is, the function
such that $\pi_i(d_1, \ldots, d_m) = d_i$ for all $(d_1, \ldots, d_m) \in D^m$.
Each function $\pi_i$ is a polymorphism of all relations,
and is hence a polymorphism of $\Gamma$;
it follows that for each $i \in \{ 1, \ldots, m \}$, the tuple
$(t_{i1}, \ldots, t_{in}) = 
(\pi(t_{11}, \ldots, t_{m1}), \ldots, \pi(t_{1n}, \ldots, t_{mn}))$
is in $R'$.
\end{pf}

Theorem~\ref{thm:galois-connection} also holds for 
infinite constraint languages; we leave the proof of this as an exercise.

\begin{exercise}
Prove Theorem~\ref{thm:galois-connection} for
infinite constraint languages 
(constraint languages containing infinitely many relations)
over finite domains
by adapting the 
given proof for Theorem~\ref{thm:galois-connection}.
\end{exercise}

We have observed that the complexity of $\csp(\Gamma)$ effectively
depends on the set $\la \Gamma \ra$, and we just showed that the set 
$\la \Gamma \ra$ can be computed from the polymorphisms of $\Gamma$.
This suggests that the polymorphisms of a constraint language $\Gamma$
can be used to derive information on the complexity of $\csp(\Gamma)$,
which we show as follows.

\begin{theorem}
\label{thm:pol-reduction}
Let $\Gamma$ and $\Gamma'$ be finite constraint languages.
If $\pol(\Gamma) \subseteq \pol(\Gamma')$, then 
$\Gamma' \subseteq \la \Gamma \ra$ and
$\csp(\Gamma')$ reduces to $\csp(\Gamma)$.
\end{theorem}

This theorem was proved by Jeavons~\cite{jeavons98}, and
can be seen as an operational
analog of Proposition~\ref{prop:expressivity-reduction}.
Note that this theorem implies that two constraint languages
$\Gamma$, $\Gamma'$ having the same polymorphisms reduce to each other,
and hence are of the same complexity.

\begin{pf}
The containment $\pol(\Gamma) \subseteq \pol(\Gamma')$ implies
the containment $\inv(\pol(\Gamma)) \supseteq \inv(\pol(\Gamma'))$.
Invoking Theorem~\ref{thm:galois-connection}, we obtain that
$\la \Gamma \ra \supseteq \la \Gamma' \ra$.  This implies that
$\Gamma' \subseteq \la \Gamma \ra$, and the conclusion follows from
Proposition~\ref{prop:expressivity-reduction}.
\end{pf}

Having established that the complexity of a constraint language $\Gamma$
is intimately linked to its set of polymorphisms $\pol(\Gamma)$,
a natural inclination at this point is to inquire as to the
structure of the sets $\pol(\Gamma)$.  
In fact, each set of operations
having this form is an instance of an algebraic object called
a \emph{clone}.

\begin{definition}
A \emph{clone} is a set of operations that
\begin{itemize}

\item contains all \emph{projections}, that is, 
the operations
$\pi_i^m: D^m \rightarrow D$ with $1 \leq i \leq m$ such that
$\pi_i^m(d_1, \ldots, d_m) = d_i$ for all $d_1, \ldots, d_m \in D$, and

\item is closed under composition, where the composition of 
an arity $n$ operation $f: D^n \rightarrow D$ and $n$ arity $m$ operations
$f_1, \ldots, f_n: D^m \rightarrow D$ is defined to be the arity $m$
operation 
$g: D^m \rightarrow D$ such that
$g(d_1, \ldots, d_m) = f(f_1(d_1, \ldots, d_m), \ldots, f_n(d_1, \ldots, d_m))$
for all $d_1, \ldots, d_m \in D$.

\end{itemize}
\end{definition}

We will say that an operation $f$ (or more generally, a set of operations $F$)
\emph{generates} an operation $g$ if every clone containing $f$
(respectively, $F$) also contains $g$.

\begin{example}
Let $f: D^n \rightarrow D$ be an operation.  The operation $f$
generates any operation $g: D^m \rightarrow D$ 
obtained by reordering and identifying arguments of $f$.
Formally, let $i: \{ 1, \ldots, n \} \rightarrow \{ 1, \ldots, m \}$
be any mapping.  
We claim that $f$ generates the operation 
$g$ such that
$g(d_1, \ldots, d_m) = f(d_{i(1)}, \ldots, d_{i(n)})$ for all
$d_1, \ldots, d_m \in D$.
This is because $g$ may be viewed as the composition of
$f$ with the projections $\pi_{i(1)}^m, \ldots, \pi_{i(n)}^m$.
\end{example}

\begin{example}
Let $f: D^2 \rightarrow D$ be a binary operation.  
We show that $f$ generates the operation $g(x,y,z) = f(x, f(y, z))$.
By the previous example, $f$ generates the operation
$f_{23}(x,y,z) = f(y,z)$.  The operation $g$ may be viewed as the
composition of $f$ with $\pi_1^3$ and $f_{23}$.
\end{example}

The computer scientist will recognize that a set of operations $F$
generates any operation $g(x_1, \ldots, x_m)$ that can be represented
as an acyclic circuit where the inputs are the variables 
$x_1, \ldots, x_m$ and the gates are operations from $F$.

\begin{prop}
For all constraint languages $\Gamma$, the set of operations
$\pol(\Gamma)$ is a clone.
\end{prop}

\begin{pf} 
Let $\Gamma$ be a constraint language, and
let $\tup{t_1}, \ldots, \tup{t_m}$ 
be elements of a relation $R$ from $\Gamma$,

For any $i$, we have
$\pi_i^m(\tup{t_1}, \ldots, \tup{t_m}) = \tup{t_i}$ which is in $R$
by assumption.  
Note that we intend that operations are applied coordinate-wise
to tuples, as in Definition~\ref{def:polymorphism}.

Now, suppose that $f: D^n \rightarrow D$ and
$f_1, \ldots, f_n: D^m \rightarrow D$ are polymorphisms of $R$,
and that $g$ is the composition of $f$ and $f_1, \ldots, f_n$.
Define $\tup{s_i} = f_i(\tup{t_1}, \ldots, \tup{t_m})$ for all
$i = 1, \ldots, n$.  Since the $f_i$ are polymorphisms of $R$,
we have that all tuples $\tup{s_i}$ are in $R$.
Since $f$ is also a polymorphism of $R$,
it follows that the tuple $\tup{s} = f(\tup{s_1}, \ldots, \tup{s_n})$
is contained in $R$.  Clearly, $g(\tup{t_1}, \ldots, \tup{t_m}) = \tup{s}$,
and we conclude that $g$ is a polymorphism of $R$.
\end{pf}

We are now ready to state Schaefer's theorem, which describes
the complexity of $\csp(\Gamma)$ for all boolean constraint languages
$\Gamma$.
Given the connection between the complexity of $\csp(\Gamma)$
and the set of polymorphisms of $\Gamma$, it should come as no surprise
that there is a description of the complexity of $\csp(\Gamma)$
in terms of polymorphisms.  Not only is this the case, but there
is a remarkably clean
description based on polymorphisms, stating that a boolean problem
$\csp(\Gamma)$
is tractable precisely when 
the constraint language $\Gamma$ has one of six polymorphisms.

Here comes the theorem statement.  
We refer to the unary operation $u_0: \{ 0, 1 \} \rightarrow \{ 0, 1 \}$ 
such that
$u_0(0) = u_0(1) = 0$ as \emph{the} constant operation $0$,
and similarly to the unary operation $u_1: \{ 0, 1 \} \rightarrow \{ 0, 1 \}$ 
such that
$u_1(0) = u_1(1) = 1$ as \emph{the} constant operation $1$.
Recall that the operation $\majority: \{ 0, 1 \}^3 \rightarrow \{ 0, 1 \}$
is defined by
$\majority(x, y, z) = (x \wedge y) \vee (x \wedge z) \vee (y \wedge z)$,
and the operation $\minority: \{ 0, 1 \}^3 \rightarrow \{ 0, 1 \}$
is defined by 
$\minority(x, y, z) = x \oplus y \oplus z$.

\begin{theorem} (Schaefer's theorem~\cite{schaefer78} -- algebraic formulation)
\label{thm:schaefer}
Let $\Gamma$ be a finite boolean constraint language.
The problem $\csp(\Gamma)$ is polynomial-time tractable
if $\Gamma$ has one of the following six operations as a polymorphism:
\begin{itemize}

\item the constant operation $0$,

\item the constant operation $1$,

\item the boolean AND operation $\wedge$,

\item the boolean OR operation $\vee$,

\item the operation $\majority$,

\item the operation $\minority$.

\end{itemize}
Otherwise, the problem $\csp(\Gamma)$ is NP-complete.
\end{theorem} 

Schaefer's theorem was originally formulated in terms of properties
of relations~\cite{schaefer78};
Jeavons~\cite{jeavons98} recognized
 that the algebraic formulation 
was possible.  Jeavons~\cite{jeavons98} also pointed out that the
algebraic formulation gives rise to a polynomial-time test for deciding
if a given constraint language satisfies one of the six tractability
conditions.  
However, the algebraic formulation not only 
makes it easy for machines to test for tractability, it makes it easy
for human beings as well!

\begin{exercise}
\label{exercise:r03-r33}
Let $\Gamma$ be the constraint language $\{ R_{0, 3}, R_{3, 3} \}$
where $R_{0, 3}$ and $R_{3, 3}$ are defined as in Example~\ref{ex:3-sat}.
Show, using Theorem~\ref{thm:schaefer}, that $\csp(\Gamma)$ is NP-complete.
\end{exercise}

\begin{exercise}
Let $\Gamma$ be the constraint language $\{ \nae \}$, where
$\nae = \{ 0, 1 \}^3 \setminus \{ (0, 0, 0), (1, 1, 1) \}$
is the relation from Example~\ref{ex:nae}.
Show, using Theorem~\ref{thm:schaefer}, that $\csp(\Gamma)$ is NP-complete.
\end{exercise}

\newcommand{\oneinthree}{R_{\mathsf{1 in 3}}}

\begin{exercise}
Let $\Gamma$ be the constraint language $\{ \oneinthree \}$, where
$\oneinthree$ denotes the ternary relation containing all tuples
with exactly one $1$, that is,
$\{ (1, 0, 0), (0, 1, 0), (0, 0, 1) \}$.
Show, using Theorem~\ref{thm:schaefer}, that $\csp(\Gamma)$ is NP-complete.
\end{exercise}

\begin{exercise}
\label{exercise:c0c1}
Let $C_0$ be the arity one relation $\{ (0) \}$, 
$C_1$ be the arity one relation $\{ (1) \}$,
and $S$ be the ternary relation 
$$\{ (a, b, c) \in \{ 0, 1 \}^3: (a = b) \vee (b = c) \}.$$
Let $\Gamma$ be the constraint language $\{ C_0, C_1, S \}$.
Show, using Theorem~\ref{thm:schaefer}, that $\csp(\Gamma)$ is NP-complete.
\end{exercise}

\section{Schaefer tractability: the good news, first}
\label{section:tractability}

In this section and the next, we establish 
Schaefer's Theorem~\ref{thm:schaefer}.  We begin with the good news
in this section by establishing that each of the six operations
listed in the theorem statement, as polymorphisms, guarantee tractability.  
We consider each of the operations in turn.  
In each case, we assume that 
$\Gamma$ is a finite constraint language having
the operation being considered as a polymorphism,
and demonstrate that $\csp(\Gamma)$ is polynomial-time tractable.

\paragraph{\bf The constant operation $0$.}
In this case, an instance of $\csp(\Gamma)$ is satisfiable
if and only if for all constraints $R(v_1, \ldots, v_k)$, the relation
$R$ is non-empty.
Clearly, this condition is necessary for satisfiability.
When the condition holds, we claim that the function mapping all variables to 
$0$ satisfies all constraints.  This is because,
in a constraint $R(v_1, \ldots, v_k)$,
 if $R$ is non-empty,
applying the polymorphism $0$ to any tuple in $R$ yields
the all-zero tuple $(0, \ldots, 0)$.

\paragraph{\bf The constant operation $1$.}
The reasoning in this case is identical to the case of the constant operation
$0$, but with the value $1$ in place of $0$.

\paragraph{\bf The boolean AND operation $\wedge$.}
In this case, we can apply a general inference algorithm
for constraint satisfaction problems called
\emph{arc consistency}.\footnote{
It should be noted that the literature contains many variants and definitions
of arc consistency.}
Assume that we have an instance of the CSP with variable set $V$
and domain $D$.
For a constraint $C = R(v_1, \ldots, v_k)$, define
$\pi_{v_i}(C) = \{ t_i: (t_1, \ldots, t_k) \in R \}$, 
for all variables $v_1, \ldots, v_k$,
and
$\pi_w(C) = D$
for all variables $w \in V \setminus \{ v_1, \ldots, v_k \}$.
Note that for any solution $f$ to the CSP, and any constraint $C$,
and variable $v$, it must hold that $f(v) \in \pi_{v}(C)$.

The following is the arc consistency algorithm.  The intuition is that,
using the sets $\pi_v(C)$, we ``tighten'' the constraints by removing
unusable tuples from the relations.  If some relation becomes empty,
we can conclude that the entire problem is unsatisfiable.

\begin{quote}
\textsc{Arc consistency algorithm}\\
Input: an instance of the CSP.\\
\begin{tabular}{rl}

1 & For each variable $v$, define $D_v$ to be $\cap_C \pi_v(C)$
where the intersection is
over all constraints $C$.\\
2 & For each constraint $R(v_1, \ldots, v_k)$, replace $R$ with
$R \cap (D_{v_1} \times \cdots \times D_{v_k})$.\\
  & If $R$ becomes empty, then terminate and report ``unsatisfiable''.\\
3 & If any relations were changed in step 2, goto step 1.
Otherwise, halt.\\
\end{tabular}
\end{quote}

Let us study this algorithm.  First, we show that if it reports
``unsatisfiable'', it does so correctly.  As we mentioned,
if $f$ is a solution to the input CSP, then for all constraints $C$ and
all variables $v$, it must hold that $f(v) \in \pi_{v}(C)$.
This implies that if $f$ is a solution to the CSP, 
we have $f(v) \in D_v$ for all variables $v$, where $D_v$ is the set computed
in step 1.  It follows that step 2 preserves the set of solutions
to the CSP: since any solution $f$ obeys $f(v) \in D_v$ for all 
variables $v$, a constraint $R(v_1, \ldots, v_k)$ is satisfied by $f$
even when $R$ is replaced with $R \cap (D_{v_1} \times \cdots \times D_{v_k})$.

Next, we show that if the algorithm halts in step 3, then there
exists a solution, assuming that each relation originally had
the $\wedge$ operation 
as a polymorphism.  Define the sets $D_v$ as in step 1 of the algorithm,
and define a mapping $f: V \rightarrow \{ 0, 1 \}$ as follows:
$$f(v)=\left\{
\begin{array}{ll}
0 & \mbox{if } D_v = \{ 0 \} \\
1 & \mbox{if } D_v = \{ 1 \} \\
0 & \mbox{if } D_v = \{ 0, 1 \} \\
\end{array}\right. $$
We claim that $f$ is a satisfying assignment.  Let $R(v_1, \ldots, v_k)$
be any constraint.  Let $\tup{s} = (s_1, \ldots, s_k)$ be the tuple obtained
by applying the $\wedge$ operation to all tuples of $R$, in any order.
That is, let $\tup{t_1}, \ldots, \tup{t_m}$ be the tuples of $R$
(in any order), and define $\tup{s}$ to be
$$((\ldots(( \tup{t_1} \wedge \tup{t_2} ) \wedge \tup{t_3}) \wedge \ldots) \wedge t_m)$$
where the $\wedge$ operator is applied to two tuples coordinate-wise,
as in Definition~\ref{def:polymorphism}.
Since the relation $R$ originally had $\wedge$ as a polymorphism and contained 
the tuples $\tup{t_1}, \ldots, \tup{t_m}$, it suffices to show that
$(f(v_1), \ldots, f(v_k)) = (s_1, \ldots, s_k)$.
Let $v_j$ be one of the variables $v_1, \ldots, v_k$, and for
each $i = 1, \ldots, m$, 
let us denote the tuple $\tup{t_i}$ by $(t_{i1}, \ldots, t_{ik})$.
If $0 \in D_{v_j}$, then there exists a tuple $\tup{t_l}$ such that
$t_{lj} = 0$, and thus $s_j = 0$.  Otherwise, $D_{v_j} = \{ 1 \}$,
and for all tuples $\tup{t_i}$ we have $t_{ij} = 1$, from which 
it follows that $s_j = 1$.

\paragraph{\bf The boolean OR operation $\vee$.}
The reasoning in this case is identical to the case of the boolean AND
operation $\wedge$, but with the roles of the values $0$ and $1$ swapped.

\paragraph{\bf The operation $\majority$.}
Let $\phi$ be a set of constraints over variable set $V$.
We say that $f: W \rightarrow D$, for $W$ a subset of $V$,
is a \emph{partial solution} of $\phi$ if for every constraint
$R(v_1, \ldots, v_k) \in \phi$, there exists a tuple
$(d_1, \ldots, d_k) \in R$ such that 
$f(v_i) = d_i$ for all $v_i \in W$.
We give an inference algorithm in the spirit of the arc consistency
algorithm given for the boolean AND operation.

\begin{quote}
\textsc{Algorithm for majority polymorphism}\\
Input: an instance $\phi$ of the CSP with variable set $V$.\\
\begin{tabular}{rl}
1 &  For each non-empty subset $W = \{ w_1, \ldots, w_l \}$
of $V$ of size $l \leq 3$,\\
  & add the constraint $D^l(w_1, \ldots, w_l)$
to $\phi$.\\
2 & For each constraint $R(w_1, \ldots, w_l)$ of $\phi$ with $l \leq 3$,
compute the set\\
  & $R' = \{ (f(w_1), \ldots, f(w_l))~|~ f:  \{ w_1, \ldots, w_l \} \rightarrow D \mbox{ is a partial solution of the instance } \phi \}.$\\
  & Then, replace $R$ with $R'$.\\
  & If $R$ becomes empty,  terminate and report ``unsatisfiable''.\\
3 & If any relations were changed in step 2, goto step 2 and repeat it.
Otherwise, halt.
\end{tabular}
\end{quote}

First, we observe that each replacement in step 2 preserves the 
set of solutions.  This is because any restriction of a solution is
a partial solution.  It follows that
if the algorithm reports ``unsatisfiable''
in step 2, it does so correctly.

We thus need to show that if the algorithm halts in step 3, then there
exists a solution, assuming that each relation originally had the
$\majority$ operation as a polymorphism.
(We assume that the CSP has three or more variables; if it has two
or fewer, it is clear that there exists a solution.)
It is straightforward to verify that each replacement of step 2
preserves all polymorphisms of the CSP, so when the algorithm halts,
the resulting CSP has the $\majority$ operation as a polymorphism.
The following definition is key.

\begin{definition}
Let $n \geq 0$.  An instance of the CSP with variable set $V$ has the 
\emph{$n$-extension property} if, given any subset $W \subseteq V$
of size $|W| = n$ and a variable $v \in V$, 
any partial solution $f: W \rightarrow D$ can be extended
to a partial solution $f': W \cup \{ v \} \rightarrow D$.
\end{definition}

Assume that the algorithm halts in step 3; 
then, 
for every pair of variables $W = \{ w, w' \}$
there exists a partial solution $f: W \rightarrow D$.
This is because there exists a constraint $S(w, w')$ with $W$
as its variables; the relation $S$ is non-empty and 
so we can define $f$ to be any assignment such that
$(f(w), f(w')) \in S$.
Moreover,
$\phi$ has the $2$-extension property:
suppose that
$W = \{ w, w' \}$ is a pair of variables, 
$v \in V$ is a variable,
and $f: W \rightarrow D$ is a partial solution.
If $v \in W$, then $f$ itself is the desired extension,
so we assume that $v \notin W$.
There exists a constraint $T(w, w', v)$ in the CSP instance;  
since $f$ is a partial solution, there exists a tuple
$(b, b', d) \in T$ such that $f(w) = b$, $f(w') = b'$.
The extension $f': \{ w, w', v \} \rightarrow D$ of 
$f$ such that $f'(v) = d$ is a partial solution.

The following theorem shows that, as a consequence of $\phi$
having the $2$-extension property, it has the $n$-extension property
for all $n \geq 3$.

\begin{theorem}
Let $\phi$ be an instance of $\csp(\Gamma)$ where $\Gamma$ has
the $\majority$ operation as a polymorphism.
If $\phi$ has the $2$-extension property, then for all $n \geq 3$
it holds that $\phi$ has the $n$-extension property.
\end{theorem}

Let $u_1, \ldots, u_m$ be an ordering of the variables $V$ in $\phi$.
We have shown that there is a partial solution on any pair of variables,
so in particular there is a partial solution
$f_2: \{ u_1, u_2 \} \rightarrow D$.
In addition, we showed that the $2$-extension property holds,
so there is a partial solution $f_3: \{ u_1, u_2, u_3 \} \rightarrow D$.
By the theorem, in an iterative manner
we can define, for all $n \geq 3$,
a partial solution
$f_{n+1}: \{ u_1, \ldots, u_{n+1} \} \rightarrow D$ extending
the partial solution
$f_n: \{ u_1, \ldots, u_n \} \rightarrow D$.
The mapping $f_m: V \rightarrow D$ is then a total solution.
Therefore, we need only prove the theorem to 
conclude our discussion of the
$\majority$ operation.

\begin{pf}
We prove this by induction.  
Suppose that $\phi$ has the $n$-extension property, with $n \geq 2$;
we show that $\phi$ has the $(n+1)$-extension property.
Let $W$ be a set of size $n + 1$, let $v$ be a variable in
$V$, and let $f: W \rightarrow D$ be a partial solution.
We want to prove that there is an extension $f': W \cup \{ v \} \rightarrow D$
of $f$ that is a partial solution.
If $v \in W$, we can take $f' = f$, so we assume that $v \notin W$.
Let $w_1, w_2, w_3$ be distinct variables in $W$,
and set $W_i = W \setminus \{ w_i \}$ for $i \in \{ 1, 2, 3 \}$.
Let $f_1$, $f_2$, $f_3$ be the restrictions of $f$ to 
$W_1$, $W_2$, and $W_3$, respectively.
Since $\phi$ has the $n$-extension property,
for each $i \in \{ 1, 2, 3 \}$ there exists an extension
$f'_i: W_i \cup \{ v \} \rightarrow D$ of $f_i$.
Define $f'$ to be the extension of $f$ with 
$f'(v) = \majority(f'_1(v), f'_2(v), f'_3(v))$.

We claim that $f'$ is a partial solution.
Let $R(v_1, \ldots, v_k)$ be any constraint.
Since each $f'_i$ is a partial solution, for each $i \in \{ 1, 2, 3 \}$
there exists a tuple $(d_1^i, \ldots, d_k^i) \in R$
such that $f'_i(v_j) = d_j^i$ when $f'_i$ is defined on $v_j$.
We show that $f'$ is a partial solution via the tuple
$(\majority(d_1^1, d_1^2, d_1^3), \ldots, \majority(d_k^1, d_k^2, d_k^3))$.
We consider each variable $v_j$, in cases.
\begin{itemize}

\item Case: $v_j = w_i$ for some $i \in \{ 1, 2, 3 \}$.
For the values $s \in \{ 1, 2, 3 \} \setminus \{ i \}$,
we have that $f_s(v_j)$ is defined and equal to 
$d_j^s = f(v_j)$.  So, two of the values
$d_j^1, d_j^2, d_j^3$ are equal to 
$f(v_j)$ and thus
$f'(v_j) = f(v_j) = \majority(d_j^1, d_j^2, d_j^3)$.

\item Case: $v_j \in W \setminus \{ w_1, w_2, w_3 \}$.
For each $i \in \{ 1, 2, 3 \}$, it holds that
$f'(v_j) = f(v_j) = f_i(v_j) = d_j^i$ and thus
$f'(v_j) = \majority(d_j^1, d_j^2, d_j^3)$.

\item Case: $v_j = v$.  We have
$f'(v) = \majority(f'_1(v), f'_2(v), f'_3(v)) = \majority( d_j^1, d_j^2, d_j^3)$.
\end{itemize}
\end{pf}

We remark that the presentation of this proof is based on the proof
of a more general result in~\cite{cd05-pebblegames}.

\paragraph{\bf The operation $\minority$.}
We begin by showing that any constraint can be 
converted into a conjunction of \emph{linear equations}.  Here,
by a linear equation we mean 
an equation of the form
$a_1 \oplus \cdots \oplus a_l = b_1 \oplus \cdots \oplus b_m$
where each $a_i$ and each $b_i$ is either a constant or a variable.

\begin{theorem}
\label{thm:minority-to-linear-equations}
Let $R \subseteq \{ 0, 1 \}^k$ be a relation having the 
$\minority$ operation as a polymorphism.  Then the constraint
$R(x_1, \ldots, x_k)$ can be expressed as the conjunction of
linear equations.
\end{theorem}

\begin{pf}
We prove this theorem by induction on the arity $k$.
The theorem is straightforward to prove for arity $k = 1$, so we
assume that $k \geq 2$.
Define the relation $R_0 \subseteq D^{k-1}$ by
$R_0(x_2, \ldots, x_k) \equiv R(0, x_2, \ldots, x_k)$
and define the relation $R_1 \subseteq D^{k-1}$ by 
$R_1(x_2, \ldots, x_k) \equiv R(1, x_2, \ldots, x_k)$.
Let $\psi_0(x_2, \ldots, x_k)$ 
and $\psi_1(x_2, \ldots, x_k)$ 
be conjunctions of equations expressing
$R_0(x_2, \ldots, x_k)$ and $R_1(x_2, \ldots, x_k)$, respectively;
these conjunctions exist by induction.

If $R_0$ is empty, 
we may express $R(x_1, \ldots, x_k)$
by $(x_1 = 1) \wedge \psi_1$.  
Similarly, if $R_1$ is empty,
we may express $R(x_1, \ldots, x_k)$
by $(x_1 = 0) \wedge \psi_0$.  
So, assume that both $R_0$ and $R_1$ are non-empty,
fix $(c^0_2, \ldots, c^0_k)$ to be a tuple in $R_0$, and 
fix $(c^1_2, \ldots, c^1_k)$ to be a tuple in $R_1$.
Define $\tup{c^0} = (0, c^0_2, \ldots, c^0_k)$ and
$\tup{c^1} = (1, c^1_2, \ldots, c^1_k)$.
Let $\tup{b} \subseteq \{ 0, 1 \}^k$ be an arbitrary tuple.
Observe that if $\tup{b} \in R$, then
$(\tup{b} \oplus \tup{c^0} \oplus \tup{c^1}) \in R$.
Moreover, if
$(\tup{b} \oplus \tup{c^0} \oplus \tup{c^1}) \in R$, then
$((\tup{b} \oplus \tup{c^0} \oplus \tup{c^1}) \oplus \tup{c^0} \oplus \tup{c^1}) = \tup{b} \in R$.
Thus, $\tup{b} \in R$ if and only if
$(\tup{b} \oplus \tup{c^0} \oplus \tup{c^1}) \in R$.
Specializing this to $b_1 = 1$, we obtain
$$(b_2, \ldots, b_k) \in R_1 \Leftrightarrow (b_2 \oplus c_2^0 \oplus c_2^1, \ldots, b_k \oplus c_k^0 \oplus c_k^1) \in R_0.$$
This implies
$$(b_1, \ldots, b_k) \in R \Leftrightarrow (b_2 \oplus c_2^0 b_1 \oplus c_2^1 b_1, \ldots, b_k \oplus c_k^0 b_1 \oplus c_k^1 b_1) \in R_0.$$
Thus, the desired equations for $R(x_1, \ldots, x_k)$ can be obtained
from the equations $\psi_0(x'_2, \ldots, x'_k)$ by substituting
$x'_i$ with $(x_i \oplus c_i^0 x_1 \oplus c_i^1 x_1)$.
\end{pf}

An instance of $\csp(\Gamma)$ can thus be reduced to the problem
of deciding if a conjunction of linear equations has a solution:
we simply replace each constraint with the equations that are given
by Theorem~\ref{thm:minority-to-linear-equations}.
The resulting conjunction of linear equations can be solved by 
the well-known Gaussian elimination algorithm.  For the sake of completeness,
we give a brief description of a polynomial-time algorithm for solving 
these linear equations.
First, we observe that it is possible to simplify linear equations
so that in each equation, each variable appears at most once: this
is because two instances of a variable in an equation can be removed
together.  Also, on each side of an equation constants may be added
together so that there is at most one constant on each side.
Next, we observe that an equation $e$ may be removed in the following way.
If it does not contain a variable, then just check to see if it is true;
if it is, then remove it, otherwise report ``unsatisfiable''.
If it does contain a variable, let us suppose for notation that
the variable is $a_1$ and the equation is
$a_1 \oplus \cdots \oplus a_l = b_1 \oplus \cdots \oplus b_m$.
The equation is equivalent to the equation
$a_1 = a_2 \oplus \cdots \oplus a_l \oplus b_1 \oplus \cdots \oplus b_m$,
which we denote by $e'$.
For each of the equations other than $e$, replace each instance of
 $a_1$ with the right-hand side of $e'$; 
then, remove $e$.  Perform this equation removal iteratively.
Note that by simplifying equations after replacements are performed,
we may ensure that the number of variable instances in each equation 
never exceeds $2n$, where
$n$ is the number of variables.

\begin{exercise}
Generalize the tractability of the operations $\wedge$ and $\vee$
by showing that any constraint language 
(over a finite domain)
having a semilattice operation
as a polymorphism is polynomial-time tractable via the given
arc consistency algorithm.
Recall that a semilattice operation is a binary operation that is
associative, commutative, and idempotent.  This tractability result
was first proved in~\cite{jcg97}.
\end{exercise}

\begin{exercise}
A \emph{near-unanimity operation} is an operation $f: D^k \rightarrow D$
of arity $k \geq 3$ such that for all elements $d, d' \in D$,
it holds that
$d = f(d', d, d, \ldots, d) = f(d, d', d, \ldots, d) = \cdots = f(d, \ldots, d, d')$.  In words, this means that if all but at most one of the inputs
to $f$ ``agree'', then the output is the agreed value.
(Observe that $\majority$ is the unique near-unanimity operation
of arity $k = 3$ over the domain $D = \{ 0, 1 \}$.)
Prove, by adapting the argument given for the $\majority$ operation,
that any constraint language $\Gamma$ 
(over a finite domain)
having a 
\emph{near-unanimity operation} as a polymorphism 
is polynomial-time tractable.
This tractability result was demonstrated in~\cite{jcc98}.
\end{exercise}

\section{Schaefer intractability}
\label{section:intractability}

In this section, we want to 
complete our proof of Schaefer's theorem by
showing that those constraint languages
not having one of the identified polymorphisms are NP-hard.
We first establish a general result on clones over the boolean domain,
showing that \emph{any} such clone either contains one of the four
non-unary operations of Schaefer's theorem, or is of a highly restricted form,
namely, only has operations that are \emph{essentially unary}.
We say that 
an operation $f: D^k \rightarrow D$ is \emph{essentially unary}
if there exists a coordinate $i \in \{ 1, \ldots, k \}$ 
and a unary operation $g: D \rightarrow D$
such that $f(d_1, \ldots, d_k) = g(d_i)$ for all values
$d_1, \ldots, d_k \in D$.

\begin{theorem} 
\label{thm:minimal-clones}
A clone over $\{ 0, 1 \}$ either contains only essentially unary operations,
or contains one of the following four operations:
\begin{itemize}

\item the boolean AND operation $\wedge$,

\item the booelan OR operation $\vee$,

\item the operation $\majority$,

\item the operation $\minority$.

\end{itemize}
\end{theorem}

Theorem~\ref{thm:minimal-clones} follows from a result proved by 
Post in 1941
which gave a description of \emph{all} clones
over the boolean domain~\cite{post41}; see~\cite{bcrv03-post} 
for a nice presentation of this theorem.
Those readers familiar with Rosenberg's theorem on 
minimal clones~\cite{rosenberg83-fivetypes}
will recognize that our proof draws on ideas present in the proof of that
result, although our proof is specialized to the boolean case.

When $f: D^k \rightarrow D$ is an operation, we will use 
$\hat{f}$ to denote the unary function defined by
$\hat{f}(d) = f(d, \ldots, d)$ for all $d \in D$.
We say that an operation $f: D^k \rightarrow D$ is \emph{idempotent}
if the operation $\hat{f}$ is the identity function.
As usual, we will use $\neg$ to denote the unary operation on $\{ 0, 1 \}$
mapping $0$ to $1$ and $1$ to $0$.

\begin{pf}
Suppose that $C$ is a clone containing an operation 
$f: \{ 0, 1 \}^k \rightarrow \{ 0, 1 \}$ that is not
essentially unary.  We will show that $f$ generates one of the four operations
given in the statement of the theorem.

We first consider the case when $\hat{f}$ is a constant operation.
Suppose that $\hat{f}$ is the constant $0$; 
the case where $\hat{f}$ is the constant $1$ is dual.
Since $f$ is not essentially unary, there exist elements
$a_1, \ldots, a_k \in \{ 0, 1 \}$ such that
$f(a_1, \ldots, a_k) = 1$.  Note that since $\hat{f} = 0$
we have $\{ a_1, \ldots, a_k \} = \{ 0, 1 \}$.
Now, we define the binary function $g$ by
$g(x_1, x_0) = f(x_{a_1}, \ldots, x_{a_k})$.
We have $g(0, 0) = g(1, 1) = 0$ and $g(1, 0) = 1$.
We consider two cases depending on the value of $g(0, 1)$.
If $g(0, 1) = 1$ we have that $g$ is the exclusive OR operation,
and thus $g$ generates the operation $\minority$: 
$\minority(x, y, z) = g(x, g(y, z))$.
If $g(0, 1) = 0$ then it can be verified that 
$x \wedge y = g(x, g(x, y))$.

We now suppose that $\hat{f}$ is not a constant operation.
It follows that $\hat{f}$ must be the identity mapping or 
the operation $\neg$.  
From $f$, we claim that we can generate an idempotent operation $f'$
that is not essentially unary.
If $f$ itself is idempotent, we take $f' = f$; if $\hat{f}$ is the operation
$\neg$, we take $f'(x_1, \ldots, x_k) = \hat{f}( f(x_1, \ldots, x_k) )$. 
We have established that the clone contains an idempotent operation
that is not essentially unary, or equivalently, an idempotent operation
that is not a projection.  
Let $g: \{ 0, 1 \}^m \rightarrow \{ 0, 1 \}$ 
be an operation of this type having minimal arity.  
We now break into cases depending on the arity $m$ of $g$.

Case $m = 2$: We cannot have $g(0, 1) \neq g(1, 0)$, 
otherwise the operation $g$ is a projection.  Thus
$g(0, 1) = g(1, 0)$.  If $g(0, 1) = g(1, 0) = 0$, we have
that $g$ is the operation $\wedge$, and if
$g(0, 1) = g(1, 0) = 1$, we have that $g$ is the operation $\vee$.

Case $m = 3$: By the minimality of the arity of $g$, if any two
arguments of $g$ are identified, we obtain a binary operation that
is a projection onto either its first or second coordinate.  
Considering the three possible ways of identifying two arguments of $g$,
we obtain the following eight possibilities:
\begin{center}
$
\begin{array}{cccccccccccccc}
 & (1) & (2)  & (3) & (4)  & (5) & (6) & (7) & (8)  \\
g(x, x, y) =  & x & x & x & x & y & y & y & y \\
g(x, y, x) =  & x & x & y & y & x & x & y & y \\
g(y, x, x) =  & x & y & x & y & x & y & x & y \\
\end{array}
$
\end{center}
In case (1), we have $g = \majority$, and in case (8), we have
$g = \minority$.  The cases (2), (3), and (5) cannot occur,
since then $f$ is a projection.  The remaining cases,
(4), (6), and (7), are symmetric; in each of them, $g$ generates
the operation $\majority$.  For example, in case (6), we have
$\majority(x, y, z) = g(x, g(x, y, z), z)$.

Case $m \geq 4$: We show that this case cannot occur.
First suppose that whenever two coordinates of $g$ are identified,
the result is a projection onto the identified coordinates.
Then, we have
$g(x_1, x_1, x_3, \ldots, x_m) = x_1$ and
$g(x_1, x_2, x_3, \ldots, x_3) = x_3$.
But, this implies that $g(0, 0, 1, \ldots, 1)$ is equal to both $0$ and
$1$, a contradiction.
So, there exist two coordinates such that, when identified,
the result is a projection onto a different coordinate.
Assume for the sake of notation that
$g(x_1, x_1, x_3, x_4, \ldots, x_m) = x_4$.
We claim that $g(x_1, x_2, x_1, x_4, \ldots, x_m) = x_4$.
Let $j$ be such that $g(x_1, x_2, x_1, x_4, \ldots, x_m) = x_j$.
This implies $g(x_1, x_1, x_1, x_4, \ldots, x_m) = x_j$.
Observe that 
$$g(x_1, x_1, x_3, x_4, \ldots, x_m) = x_4 \mbox{ implies } g(x_1, x_1, x_1, x_4, \ldots, x_m) = x_4$$ 
and hence $x_j = x_4$.
We can similarly show that 
$g(x_1, x_2, x_2, x_4, \ldots, x_m) = x_4$.
For any assignment to the variables $x_1, \ldots, x_m$,
one of the equalities $x_1 = x_2$, $x_1 = x_3$, $x_2 = x_3$
must hold, since our domain $\{ 0, 1 \}$ has two elements,
and thus $g$ is a projection.
\end{pf}

Next, we establish a lemma on which all of our intractability results
will be based.
Let us say that an essentially unary operation $f: D^k \rightarrow D$ 
\emph{acts as a permutation} if there exists a coordinate
$i \in \{ 1, \ldots, k \}$ and a bijective operation
$\pi: D \rightarrow D$  such that
$f(x_1, \ldots, x_k) = \pi(x_i)$.

\begin{lemma}
\label{lemma:permutations}
If $\Gamma$ is a finite boolean constraint language such that
$\pol(\Gamma)$ contains only essentially unary operations that act
as permutations, then for any finite boolean constraint language $\Gamma'$,
it holds that $\csp(\Gamma')$ reduces to $\csp(\Gamma)$.
\end{lemma}

\begin{pf}
If $\pol(\Gamma)$ contains only projections, then 
$\pol(\Gamma) \subseteq \pol(\Gamma')$ and we can apply 
Theorem~\ref{thm:pol-reduction}.

Otherwise, we define a constraint language $\Gamma''$ that contains,
for every relation $R' \in \Gamma'$, a relation $R''$ defined by
$R'' = \{ (0, t_1, \ldots, t_k): (t_1, \ldots, t_k) \in R' \} \cup 
\{ (1, \neg t_1, \ldots, \neg t_k): (t_1, \ldots, t_k) \in R' \}$.
The constraint language $\Gamma''$ has the operation $\neg$ as a polymorphism,
and hence, by the assumption, all operations in $\pol(\Gamma)$ 
as a polymorphism.
Thus, by Theorem~\ref{thm:pol-reduction}, 
$\csp(\Gamma'')$ reduces to $\csp(\Gamma)$, and it
suffices to show that $\csp(\Gamma')$ reduces to $\csp(\Gamma'')$.

Given an instance $\phi'$ of $\csp(\Gamma')$ having variables $V$, 
we create an instance $\phi''$ of
$\csp(\Gamma'')$ as follows.
Introduce a fresh variable
$b_0 \notin V$, 
and for every constraint $R'(v_1, \ldots, v_k)$ occurring in our
instance $\phi'$, we create a constraint
$R''(b_0, v_1, \ldots, v_k)$ in the instance $\phi''$.
Suppose that $f: V \rightarrow \{ 0, 1 \}$ satisfies $\phi'$;
 then, the extension
of $f$ mapping $b_0$ to $0$ satisfies $\phi''$.
Suppose that $g: V \cup \{ b_0 \} \rightarrow \{ 0, 1 \}$ satisfies $\phi''$;
then, if $g(b_0) = 0$, the restriction of $g$ to $V$ satisfies $\phi'$,
and if $g(b_0) = 1$, the mapping $f: V \rightarrow \{ 0, 1 \}$
defined by $f(v) = \neg g(v)$ for all $v \in V$, satisfies $\phi'$.
\end{pf}

We can now prove Schaefer's Theorem.

\begin{pf} (Theorem~\ref{thm:schaefer})
Let $\Gamma$ be a finite boolean constraint language.  
If $\Gamma$ has one of the six operations given in the theorem statement
as a polymorphism, then, as discussed in Section~\ref{section:tractability},
the problem $\csp(\Gamma)$ is polynomial-time tractable.
Otherwise, since $\Gamma$ does not have any of the operations
$\{ \wedge, \vee, \majority, \minority \}$ as a polymorphism, 
by Theorem~\ref{thm:minimal-clones}, 
the clone $\pol(\Gamma)$ contains only essentially unary operations.
The only unary operations on $\{ 0, 1 \}$ are the two constant 
operations and the two permutations (the identity and $\neg$).
Since $\pol(\Gamma)$ does not contain either of the two constant operations,
the set 
$\pol(\Gamma)$ contains only essentially unary operations that act
as permutations, and the result follows from
Lemma~\ref{lemma:permutations}
by taking $\Gamma'$ to be a finite boolean constraint language where
$\csp(\Gamma')$ is known to be NP-hard.
For instance, we can take $\Gamma'$ to be the constraint language
identified by Proposition~\ref{prop:3-sat-np-hard}.
\end{pf}

We close the section with some exercises.

\begin{exercise}
Using Theorem~\ref{thm:minimal-clones}, observe that 
for each of the constraint languages $\Gamma$ in 
Exercises~\ref{exercise:r03-r33}-\ref{exercise:c0c1},
the only polymorphisms of $\Gamma$ are the projections.
Then, using 
Theorem~\ref{thm:galois-connection},
observe that $\la \Gamma \ra$ is the set of all relations.
\end{exercise}

\begin{exercise}
Prove that
if $\Gamma$ is a finite constraint language (over a finite domain $D$),
then there exists a finite constraint language $\Gamma'$ 
(over a finite domain) such that $\csp(\Gamma)$ reduces to and from
$\csp(\Gamma')$, and all essentially unary polymorphisms of
$\Gamma$ act as permutations.  
This result was established in~\cite{bjk05}.
{\bf Hint:} let $f: D \rightarrow D$ be a unary polymorphism of
$\Gamma$ having minimal image size, and consider $\Gamma' = f(\Gamma)$.
\end{exercise}

The following generalization of Lemma~\ref{lemma:permutations} is 
known~\cite{bjk05}.

\begin{lemma} (follows from~\cite{bjk05})
\label{lemma:bjk-permutations}
If $\Gamma$ is a finite constraint language (over a finite domain $D$)
whose essentially unary polymorphisms all act as permutations,
then the problem $\csp(\Gamma)$ reduces to and from $\csp(\Gamma_c)$,
where $\Gamma_c = \Gamma \cup \{ \{ (d) \}: d \in D \}$.
That is, $\Gamma_c$ is the constraint language obtained by
augmenting $\Gamma$ with all unary relations of size one.
\end{lemma}

\begin{exercise}
Derive Lemma~\ref{lemma:permutations} from Lemma~\ref{lemma:bjk-permutations}.
\end{exercise}

\begin{exercise}
Prove Lemma~\ref{lemma:bjk-permutations}.  
{\bf Hint:} construct an instance of $\csp(\Gamma)$ with variable set $D$
whose solutions $f: D \rightarrow D$ are the unary polymorphisms of $\Gamma$.
\end{exercise}

\section{Adding quantification to the mix}
\label{section:quantified}

In this section, we consider \emph{quantified} constraint satisfaction
problems.  We prove a classification result for the boolean domain
analogous to Schaefer's theorem.  The problems we study are defined
as follows.

\begin{definition}
Let $\Gamma$ be a finite constraint language.  The problem $\qcsp(\Gamma)$
is to decide the truth of a formula of the form
$Q_1 v_1 \ldots Q_n v_n \cons$, where 
\begin{itemize}
\item each $Q_i \in \{ \forall, \exists \}$  is a quantifier,
\item each $v_i$ is a variable, and
\item $\cons$ is the finite conjunction of constraints having relations from $\Gamma$ and variables from $\{ v_1, \ldots, v_n \}$.
\end{itemize}
The sequence $Q_1 v_1 \ldots Q_n v_n$ is called the \emph{quantifier prefix} of the formula.
\end{definition}

For each constraint language $\Gamma$, the problem
$\qcsp(\Gamma)$ is a generalization of $\csp(\Gamma)$:
the problem $\csp(\Gamma)$ can be viewed
as the restriction of $\qcsp(\Gamma)$ to instances having only
existential quantifiers.
It is well-known that each problem $\qcsp(\Gamma)$ belongs to 
the complexity class PSPACE; it is also known that there exist
constraint languages $\Gamma$ such that $\qcsp(\Gamma)$ is 
PSPACE-hard.

\begin{prop} \cite{sm73}
\label{prop:3-sat-pspace-hard}
The problem $\qcsp(\Gamma_3)$, where $\Gamma_3$ as is defined in 
Example~\ref{ex:3-sat}, is PSPACE-hard.
\end{prop}

This section will prove the following theorem, which classifies
every problem of the form $\qcsp(\Gamma)$ as tractable or PSPACE-complete.

\begin{theorem}
\label{thm:qcsp-classification}
Let $\Gamma$ be a finite boolean constraint language.
The problem $\qcsp(\Gamma)$ is polynomial-time tractable if $\Gamma$ has
one of the four operations $\{ \wedge, \vee, \majority, \minority \}$
as a polymorphism; otherwise, the problem $\qcsp(\Gamma)$ is 
PSPACE-complete.
\end{theorem}
Theorem~\ref{thm:qcsp-classification} is a 
``wide dichotomy theorem'': it shows that only the complexity classes
P and PSPACE---and none of the complexity classes inbetween---are
characterized by the boolean problems $\qcsp(\Gamma)$.
Also note that, in contrast to Schaefer's theorem, here the
constant polymorphisms are not strong enough to guarantee
tractability of $\qcsp(\Gamma)$;
however, see Exercise~\ref{exercise:constant-polymorphisms-in-qcsp}
below for a related result.

A full proof of
Theorem~\ref{thm:qcsp-classification} was given by 
Creignou et al.~\cite{cks01},
although a partial classification 
was claimed by Schaefer~\cite{schaefer78}.
Related work by Dalmau~\cite{dalmau97} 
established the full classification 
given by this theorem, assuming Schaefer's unproved partial classification.

As for the CSP, we can give a notion of definability for the QCSP
that allows a constraint language to ``simulate'' further relations.
Whereas for the CSP we considered pp-definability, here we consider
a more general notion of definability where universal quantification
is permitted.

\begin{definition}
We say that a relation $R \subseteq D^k$ is \emph{few-definable}\footnote
{
Where does the name \emph{few}-definable come from?  The LaTeX
commands for the symbols $\forall$, $\exists$, and $\wedge$.
}
from
a constraint language $\Gamma$ if for some $m \geq 0$
there exists a finite conjunction $\cons$ consisting of constraints 
and equalities $(u = v)$
over variables $\{ v_1, \ldots, v_k, x_1, \ldots, x_m \}$ 
and quantifiers $Q_1, \ldots, Q_m \in \{ \forall, \exists \}$
such that
$$R(v_1, \ldots, v_k) \equiv Q_1 x_1 \ldots Q_m x_m \cons.$$
We use $[\Gamma]$ to denote the set of all relations that are
few-definable from $\Gamma$.
\end{definition}

Note that a pp-definition is a special case of a few-definition,
and so a relation $R$ that is pp-definable from a constraint language
$\Gamma$ is also few-definable from $\Gamma$, that is, the containment
$\la \Gamma \ra \subseteq [\Gamma]$ holds.

\begin{prop}
\label{prop:qcsp-reduce}
Let $\Gamma$ and $\Gamma'$ be finite constraint languages.
If $\Gamma' \subseteq [\Gamma]$, then 
$\qcsp(\Gamma')$ reduces to $\qcsp(\Gamma)$.  
\end{prop}

\begin{pf}
The proof resembles the proof of Proposition~\ref{prop:expressivity-reduction}.
From an instance $\Phi = \prefix \phi$ of $\qcsp(\Gamma')$, 
we create an instance of $\qcsp(\Gamma)$ in the following way.
We loop over each constraint $C = R(v_1, \ldots, v_k)$ in 
the original instance,
performing the following operations for each: 
let 
$Q_1 x_1 \ldots Q_m x_m \cons$ be a few-definition of $C$ over $\Gamma$,
rename the quantified variables $x_1, \ldots, x_m$ if necessary
so that they are distinct from all other variables in the formula,
replace $C$ with $\cons$ in $\phi$, 
and add $Q_1 x_1 \ldots Q_m x_m$ to the
end of the quantifier prefix $\prefix$.
The resulting quantified formula may contain equalities
$(u = v)$; we can process such equalities one by one as follows.
Suppose that an equality $(u = v)$ is present and that $u$ comes
before $v$ in the quantifier prefix.  If $v$ is universally quantified,
then the formula is false.  Otherwise, 
remove $v$ from the quantifier prefix, and
replace all instances of $v$ with $u$.
\end{pf}

We now give a theorem that will be used to establish
the needed hardness results in the classification.

\begin{theorem}
\label{thm:qcsp-hardness-of-eu}
If $\Gamma$ is a finite boolean constraint language such that
$\pol(\Gamma)$ contains only essentially unary operations,
then for any finite boolean constraint language $\Gamma_0$,
it holds that $\qcsp(\Gamma_0)$ reduces to $\qcsp(\Gamma)$.
\end{theorem}

\begin{pf}
We prove this theorem in a sequence of steps.

\paragraph{\bf Step 1.}
We define a constraint language $\Gamma'$ that contains,
for each relation $R \in \Gamma_0$, 
the relation $R' = (R \times \{ 0, 1 \}) \cup \{ (0, \ldots, 0), (1, \ldots, 1) \}$.
Observe that $\Gamma'$ has both constant polymorphisms.
In addition, for each relation $R \in \Gamma_0$, it holds that
$R(v_1, \ldots, v_k) = \forall y (R'(v_1, \ldots, v_k, y))$
and thus $R \in [\Gamma']$.
Thus, by Proposition~\ref{prop:qcsp-reduce},
we have that $\qcsp(\Gamma_0)$ reduces to $\qcsp(\Gamma')$.

\paragraph{\bf Step 2.}
In this step, we show that $\qcsp(\Gamma')$ reduces to $\qcsp(\Gamma'')$
for a constraint language $\Gamma''$ having all unary operations
(over $\{ 0, 1 \}$) as polymorphisms.  
We proceed as in the proof of Lemma~\ref{lemma:permutations}.
For each relation $R' \in \Gamma'$, there is a relation $R'' \in \Gamma''$
defined by
$R'' = \{ (0, t_1, \ldots, t_k): (t_1, \ldots, t_k) \in R' \} \cup 
\{ (1, \neg t_1, \ldots, \neg t_k): (t_1, \ldots, t_k) \in R' \}$.
Using the fact that $\Gamma'$ has the constant $0$ operation as a polymorphism,
it can be seen that $\Gamma''$ has all unary operations as polymorphisms.
Given an instance $\Phi'$ of $\qcsp(\Gamma')$, we create an instance of
$\qcsp(\Gamma'')$ as follows.  We introduce a new variable $b_0$,
and for every constraint $R'(v_1, \ldots, v_k)$ appearing in our
instance $\Phi'$, we create a constraint
$R''(b_0, v_1, \ldots, v_k)$ in the instance $\Phi''$.
The quantifier prefix of $\Phi''$ is $Q b_0$ for either quantifier
$Q \in \{ \forall, \exists \}$ followed by the quantifier prefix of $\Phi'$;
that is, the quantifier prefix of $\Phi''$ is obtained by adding,
to the quantifier prefix of $\Phi'$,  $b_0$
as the outermost quantified variable.
It is straightforward to show that $\Phi'$ is true if and only if
$\Phi''$ is true.

\paragraph{\bf Step 3.} 
Since $\Gamma''$ has all unary operations as polymorphisms,
it has all essentially unary operations as polymorphisms, and
we have $\pol(\Gamma'') \supseteq \pol(\Gamma)$.
By Theorem~\ref{thm:pol-reduction},
we have the containment $\Gamma'' \subseteq \la \Gamma \ra$,
and thus $\qcsp(\Gamma'')$ reduces to $\qcsp(\Gamma)$
by Proposition~\ref{prop:qcsp-reduce}.
\end{pf}

We can now prove the classification theorem.

\begin{pf} (Theorem~\ref{thm:qcsp-classification})
If $\Gamma$ has one of the given operations as a polymorphism, then 
the problem $\qcsp(\Gamma)$ is tractable;
we refer the reader to the original proofs~\cite{apt79,kbs87,bkf95,cks01}
and to~\cite{chen06-collapsibility} for an algebraic approach.
If $\Gamma$ does not have one of the four given operations as a polymorphism,
then by Theorem~\ref{thm:minimal-clones},
the set $\pol(\Gamma)$ contains only essentially unary operations,
and $\qcsp(\Gamma)$ is PSPACE-hard by 
Proposition~\ref{prop:3-sat-pspace-hard} and
Theorem~\ref{thm:qcsp-hardness-of-eu}.
\end{pf}

\begin{exercise}
\label{exercise:surjective-preservation}
When $\Gamma$ is a constraint language,
let $\spol(\Gamma)$ denote the set 
containing all \emph{surjective} polymorphisms of $\Gamma$.
Prove the following result, which was established by 
B\"{o}rner et al.~\cite{bbkj03}.
\begin{theoremnonumber}
Let $\Gamma$ be a constraint language over a finite domain $D$.
It holds that $[\Gamma] = \inv(\spol(\Gamma))$.
\end{theoremnonumber}
\noindent
{\bf Hint:} to show that a relation $R \in \inv(\spol(\Gamma))$
is in $[\Gamma]$, define $R'$ to be
$$R' = \{ f(\tup{t_1}, \ldots, \tup{t_m}): f \in \pol(\Gamma), \tup{t_i} \in R \times D^d \}$$
which is the smallest relation in $\la \Gamma \ra$ containing
$R \times D^d$, and consider 
$$\forall y_1 \ldots \forall y_d (R'(v_1, \ldots, v_k, y_1, \ldots, y_d)).$$
Here, we use $d$ to denote the size of $D$.
\end{exercise}

\begin{exercise}
\label{exercise:constant-polymorphisms-in-qcsp}
Let $\Gamma$ be a finite boolean constraint language.
Prove that if $[\Gamma]$ has one of the constant operations $0$, $1$
as a polymorphism, then $\qcsp(\Gamma)$ is polynomial-time tractable.
\end{exercise}

\section{Bounded alternation}
\label{section:bd-alternation}

In this section, we study the problems $\qcsp(\Gamma)$ under restricted
\emph{prefix classes}.  What do we mean by a prefix class?
For each instance of the QCSP, we may take the quantifier prefix,
eliminate the variables, and then group together consecutive 
quantifiers that are identical to obtain a pattern.
For example, from the quantifier prefix
$\exists w_1 \exists w_2 \exists w_3 \forall y_1 \forall y_2 \exists x_1 \exists x_2 \exists x_3 \exists x_4$ we obtain the pattern $\exists \forall \exists$.
An instance is said to have prefix class
$\Sigma_1$ if its pattern is $\exists$, 
prefix class
$\Pi_1$ if its pattern is $\forall$, 
prefix class
$\Sigma_2$ if its pattern is $\exists \forall$, 
prefix class
$\Pi_2$ if its pattern is $\forall \exists$, and so forth.
The collection of formulas having prefix class $\Sigma_k$ or $\Pi_k$
for some $k \geq 1$ are said to be of \emph{bounded alternation}
because in their quantifier prefixes, the number of alternations between the 
two different quantifiers is bounded above by a constant.
For all $k \geq 1$ and finite constraint languages $\Gamma$,
we define $\Sigma_k$-$\qcsp(\Gamma)$ ($\Pi_k$-$\qcsp(\Gamma)$)
to be the restriction of $\qcsp(\Gamma)$ to instances having 
prefix classes $\Sigma_k$ (respectively, $\Pi_k$).
We study the prefix classes where the innermost quantifier is existential,
that is, the classes $\Sigma_k$ for odd $k$ and $\Pi_k$ for even $k$.
It is known that the corresponding problems characterize the 
complexity classes $\Sigma_k^p$, $\Pi_k^p$ of the polynomial hierarchy.

\begin{prop} \cite{stockmeyer76,wrathall76}
\label{prop:bounded-hard}
Let $\Gamma$ be a finite constraint language over a finite domain.
For even $k \geq 2$, the problem $\Pi_k$-$\qcsp(\Gamma)$ is
in $\Pi_k^p$, and for odd $k \geq 3$, the problem
$\Sigma_k$-$\qcsp(\Gamma)$ is in $\Sigma_k^p$.
In addition, letting $\Gamma_3$ denote the constraint language
from Example~\ref{ex:3-sat}, for even $k \geq 2$,
the problem $\Pi_k$-$\qcsp(\Gamma_3)$ is $\Pi_k^p$-complete, and
the problem $\Sigma_k$-$\qcsp(\Gamma_3)$ is $\Sigma_k^p$-complete.
\end{prop}

This section proves the following classification theorem.

\begin{theorem}
\label{thm:bd-alt-classification}
Let $\Gamma$ be a finite boolean constraint language.
For all $k \geq 1$,
the problems $\Pi_k$-$\qcsp(\Gamma)$ and $\Sigma_k$-$\qcsp(\Gamma)$
are polynomial-time tractable if
$\Gamma$ has
one of the four operations $\{ \wedge, \vee,  \majority, \minority \}$
as a polymorphism; otherwise,
\begin{itemize}

\item the problem $\Pi_k$-$\qcsp(\Gamma)$ is $\Pi^p_k$-complete
 for even $k \geq 2$,
and

\item the problem $\Sigma_k$-$\qcsp(\Gamma)$ is $\Sigma^p_k$-complete for
odd $k \geq 3$.

\end{itemize}
\end{theorem}
Theorem~\ref{thm:bd-alt-classification} 
can be taken as a fine version of
the last section's classification theorem
(Theorem~\ref{thm:qcsp-classification}), showing that the 
constraint languages shown to be hard in that theorem are also hard
under bounded alternation.
Theorem~\ref{thm:bd-alt-classification} has been previously established
by Hemaspaandra~\cite{hemaspaandra04} and Bauland et al.~\cite{bbrcsv05}.

How can we prove this theorem?
The tractable cases clearly follow from 
those of the last section's classification theorem,
so we are left with proving the hardness results.
What happens if we try to establish these hardness results by 
imitating the proof of 
the hardness results of the last section
(that is, the proof of Theorem~\ref{thm:qcsp-hardness-of-eu})?
We become faced with a difficulty: this proof relied on 
Proposition~\ref{prop:qcsp-reduce},
which showed that $\qcsp(\Gamma')$ reduces to $\qcsp(\Gamma)$,
assuming $\Gamma' \subseteq [\Gamma]$.  However, the given proof
of this proposition does \emph{not} preserve the prefix class of the 
original formula: an instance $\Phi$ of
$\qcsp(\Gamma')$ is transformed into an instance of $\qcsp(\Gamma)$
by taking few-definitions for the constraints in $\Phi$ and
appending these few-definitions to the end of the quantifier prefix of $\Phi$!
To overcome this difficulty, we establish a version of 
Proposition~\ref{prop:qcsp-reduce} which \emph{does} preserve
the prefix class.

\begin{theorem}
\label{thm:bd-qcsp-reduce}
Let $\Gamma$ and $\Gamma'$ be finite constraint languages
over the same finite domain.
If $\Gamma' \subseteq [\Gamma]$, then 
\begin{itemize}

\item for even $k \geq 2$, 
there is a reduction from $\Pi_k$-$\qcsp(\Gamma')$ to $\Pi_k$-$\qcsp(\Gamma)$,
and 

\item for odd $k \geq 3$,
there is a reduction from $\Sigma_k$-$\qcsp(\Gamma')$ to
$\Sigma_k$-$\qcsp(\Gamma)$.
\end{itemize}
\end{theorem}

To the best of my knowledge, this theorem is new.  
Although here we only apply this theorem to the case of a boolean domain,
I would like to emphasize that the theorem gives a general tool
that applies for all finite domains, which I believe will be useful
for studying the bounded alternation QCSP over domains of larger size.

In order to prove Theorem~\ref{thm:bd-qcsp-reduce}, 
we establish a lemma that shows that
the relations in $[\Gamma]$ can be expressed in a particular way.

\begin{definition}
Let $R \subseteq D^k$ be a relation of arity $k$ over a finite domain $D$
of size $d = |D|$.
Say that $R$ is \emph{spread-expressed} by $R' \subseteq D^{k + d}$
if the following two properties hold:
\begin{itemize}

\item (monotonicity) 
if $b_1, \ldots, b_d, b'_1, \ldots, b'_d \in D$
are elements such that 
$\{ b_1, \ldots, b_d \} \supseteq \{ b'_1, \ldots, b'_d \}$, then for all
$(a_1, \ldots, a_k) \in D^k$,
$R'(a_1, \ldots, a_k, b_1, \ldots, b_d)$ implies
$R'(a_1, \ldots, a_k, b'_1, \ldots, b'_d)$.

\item (expression) if 
$b_1, \ldots, b_d \in D$ are elements such that
$\{ b_1, \ldots, b_d \} = D$, then 
for all $(a_1, \ldots, a_k) \in D^k$,
$R(a_1, \ldots, a_k)$ if and only if
$R'(a_1, \ldots, a_k, b_1, \ldots, b_d)$.

\end{itemize}
\end{definition}

\begin{lemma}
\label{lemma:spread-expressed}
Let $\Gamma$ be a constraint language over a finite domain $D$.
For every relation $R \in [\Gamma]$, there exists a relation
$R' \in \la \Gamma \ra$ such that $R$ is spread-expressed by $R'$.
\end{lemma}

\begin{pf}
We let $d$ denote the size of $D$ and $k$ denote the arity of $R$.
If $R$ is pp-definable from $\Gamma$ without the use of quantifiers,
then let $\phi(v_1, \ldots, v_k)$ denote such a definition of
$R(v_1, \ldots, v_k)$.  We define $R'$ by 
$R'(v_1, \ldots, v_k, y_1, \ldots, y_d) = \phi(v_1, \ldots, v_k)$.

Now suppose that the lemma is true for a relation $R$.
We want to show that it is true for a relation obtained from $R$
by ``quantifying away'' a coordinate of $R$.
Suppose that $R_2(x_1, \ldots, x_{k-1}) = Q x_k R(x_1, \ldots, x_k)$,
with $Q \in \{ \forall, \exists \}$.
We consider two cases depending on the quantifier $Q$.
In both cases, we use
$R'$ to denote a relation such that $R$ is spread-expressed
by $R'$.

Case $Q = \exists$: 
It is straightforward to verify that $R_2$ is spread-expressed by
the relation $R'_2$ defined by
$$R'_2(x_1, \ldots, x_{k-1}, y_1, \ldots, y_d) = \exists x_k R'(x_1, \ldots, x_k, y_1, \ldots, y_d).$$

Case $Q = \forall$: 
We claim that $R_2$ is spread-expressed by the relation $R'_2$ 
defined by
$$R'_2(x_1, \ldots, x_{k-1}, y_1, \ldots, y_d) = 
\bigwedge_{i=1}^d R'(x_1, \ldots, x_{k-1}, y_i, y_1, \ldots, y_d).$$
We verify this as follows.

First, we verify monotonicity.  Suppose that
$b_1, \ldots, b_d, b'_1, \ldots, b'_d \in D$
are elements such that the containment
$\{ b_1, \ldots, b_d \} \supseteq \{ b'_1, \ldots, b'_d \}$ holds.
Let $(a_1, \ldots, a_{k-1}) \in D^{k-1}$ be a tuple, and 
suppose that $R'_2(a_1, \ldots, a_{k-1}, b_1, \ldots, b_d)$ holds.
Then, by the definition of $R'_2$, we have 
$$\bigwedge_{b \in \{ b_1, \ldots, b_d \}} R'(a_1, \ldots, a_{k-1}, b, b_1, \ldots, b_d), \mbox{ implying }
\bigwedge_{b \in \{ b'_1, \ldots, b'_d \}} R'(a_1, \ldots, a_{k-1}, b, b_1, \ldots, b_d).$$ 
By the monotonicity of $R'$, this in turn implies that 
$\bigwedge_{b \in \{ b'_1, \ldots, b'_d \}} R'(a_1, \ldots, a_{k-1}, b, b'_1, \ldots, b'_d)$ holds, which is equivalent to 
$R'_2(a_1, \ldots, a_{k-1}, b'_1, \ldots, b'_d)$.

Next, suppose that
$b_1, \ldots, b_d \in D$ are elements such that
$\{ b_1, \ldots, b_d \} = D$, and let
$(a_1, \ldots, a_{k-1}) \in D^{k-1}$ be a tuple.
We have 
$$R_2(a_1, \ldots, a_{k-1}) \Leftrightarrow$$
$$\forall b \in D (R(a_1, \ldots, a_{k-1}, b)) \Leftrightarrow$$
$$\forall b \in D (R'(a_1, \ldots, a_{k-1}, b, b_1, \ldots, b_d)) \Leftrightarrow$$
$$R'_2(a_1, \ldots, a_{k-1}, b_1, \ldots, b_d).$$
\end{pf}

We now give the proof of Theorem~\ref{thm:bd-qcsp-reduce}.

\begin{pf} (Theorem~\ref{thm:bd-qcsp-reduce})
For each relation $R' \in \Gamma'$, there exists a relation
$R'' \in \la \Gamma \ra$ such that $R'$ is spread-expressed by $R''$.
Let $\Phi'$ be an instance of 
$\Pi_k$-$\qcsp(\Gamma')$ or $\Sigma_k$-$\qcsp(\Gamma')$
(with $k$ as described in the statement of the theorem).
Denote the quantifier prefix of $\Phi'$ by $\prefix'$
and the conjunction of constraints of $\Phi'$ by $\phi'$, so that
$\Phi' = \prefix' \phi'$.  
We create an 
instance $\Phi''$ of
$\Pi_k$-$\qcsp(\Gamma)$ or $\Sigma_k$-$\qcsp(\Gamma)$ 
as follows.
First, define $\prefix''$ to be a quantifier prefix
obtained from 
$\prefix'$ by introducing new universally quantified variables
$y_1, \ldots, y_d$ and placing them next to any group of universally
quantified variables in $\prefix'$, so that the prefix class is preserved.
Next, define $\phi''$ to be the conjunction of constraints
containing a constraint $R''(v_1, \ldots, y_k, y_1, \ldots, y_d)$
for every constraint $R'(v_1, \ldots, v_k)$ in $\phi'$.
The output of the reduction is the formula $\Phi'' = \prefix'' \phi''$,
but where every constraint is replaced with a pp-formula
over $\Gamma$, as in the proof of Proposition~\ref{prop:qcsp-reduce}.

To verify the correctness of the reduction, we need to 
show that $\Phi' = \prefix' \phi'$ is true if and only if
$\Phi'' = \prefix'' \phi''$ is true.  This follows from the following
cycle of implications.
Fix $g_0: \{ y_1, \ldots, y_d \} \rightarrow D$ to be a surjective mapping.
$$\prefix' \phi' \Rightarrow
\prefix' \forall y_1 \ldots \forall y_d \phi'' \Rightarrow
\prefix'' \phi'' \Rightarrow
\forall y_1 \ldots \forall y_d \prefix' \phi'' \Rightarrow
\prefix' \phi'' \mbox{ is true under } g_0 \Rightarrow
\prefix' \phi'$$
\end{pf}

We can now prove the classification theorem.

\begin{pf} (Theorem~\ref{thm:bd-alt-classification})
Following the discussion earlier in this section,
we need to show that, when $P = \Pi_k$ for an even $k \geq 2$
or $P = \Sigma_k$ for an odd $k \geq 3$, the problem
$P$-$\qcsp(\Gamma)$ is hard for the corresponding class of the polynomial
hierarchy, assuming that $\Gamma$ does not have one of the four given
polymorphisms.
By Theorem~\ref{thm:minimal-clones}, all polymorphisms of $\Gamma$ 
are
essentially unary operations.
Let $\Gamma_0$ be the finite boolean constraint language of
Proposition~\ref{prop:bounded-hard}, 
so that $P$-$\qcsp(\Gamma_0)$ is hard for the complexity class
for which we want to prove hardness.

We now follow the proof of Theorem~\ref{thm:qcsp-hardness-of-eu}.
The main modification we need to make is in step 1.
Step 1 defines a constraint language $\Gamma'$ having both constant
polymorphisms such that $\Gamma_0 \subseteq [\Gamma']$.
Here, we appeal to Theorem~\ref{thm:bd-qcsp-reduce}
(instead of Proposition~\ref{prop:qcsp-reduce})
to obtain a reduction from $P$-$\qcsp(\Gamma_0)$
to $P$-$\qcsp(\Gamma')$.
Step 2 can be carried out in a way that preserves the prefix class
of the formula, since the new variable $b_0$ can be quantified
using either quantifier (as noted); this gives us a reduction
from $P$-$\qcsp(\Gamma')$ to $P$-$\qcsp(\Gamma'')$ for a constraint language
$\Gamma''$ having all unary operations as polymorphisms.
Finally, in step 3, we have $\pol(\Gamma'') \supseteq \pol(\Gamma)$
and hence $\Gamma'' \subseteq \la \Gamma \ra$, that is,
every relation in $\Gamma''$
has a pp-definition in $\Gamma$.
Since the innermost quantifier of $P$ is existential, it follows that
$P$-$\qcsp(\Gamma'')$ can be reduced to 
$P$-$\qcsp(\Gamma)$.
\end{pf}

\begin{exercise}
Describe the complexity of all finite constraint languages
in the problems $\Pi_k$-$\qcsp(\Gamma)$ for all odd $k \geq 1$,
and the problems $\Sigma_k$-$\qcsp(\Gamma)$ for all even $k \geq 2$.
\end{exercise}

\begin{exercise}
Give an alternative, algebraic proof of
Lemma~\ref{lemma:spread-expressed}
by considering the hint of Exercise~\ref{exercise:surjective-preservation}.
\end{exercise}

\section{To infinity... and beyond?}
\label{section:to-infinity}

In recent joint work,
Manuel Bodirsky and I considered the 
quantified CSP over domains of \emph{infinite} size~\cite{bc06}.
One  class of constraint languages that we have considered
is the class of \emph{equality constraint languages}.
We say that a constraint language over domain $D$ is
an \emph{equality constraint language} if each relation is 
\emph{equality definable}, by which we mean definable
using equalities $(u = v)$ and the usual boolean connectives
$\neg$, $\wedge$, and $\vee$.  As an example, consider the 
ternary relation
$R \subseteq D^3$ defined by
$$R(x, y, z) \equiv (\neg (x = y)) \vee (y = z).$$
Clearly, the disequality relation $\neq$ is another example of an 
equality definable relation:
$$(x \neq y) \equiv (\neg (x = y)).$$
As an intuition pump, consider $\csp( \{ \neq \} )$, the 
constraint satisfaction problem over the disequality relation.
The problem $\csp( \{ \neq \} )$ is
the $|D|$-colorability problem: given a set of pairs of variables,
decide if the variables can be colored with elements from $D$ such that
each pair has different colors.  
Over a \emph{finite} domain,
the problem $\csp( \{ \neq \} )$ is of course known to be NP-complete
for $|D| \geq 3$.
On the other hand, over an \emph{infinite} domain,
the problem $\csp( \{ \neq \} )$ is trivial: 
if an instance contains a constraint of the form $v \neq v$,
it is not satisfiable, otherwise it is
satisfiable by the assignment sending all variables to different values!

Within the class of equality constraint languages, we were able to 
establish a complexity upper bound.
Let us define a
\emph{positive constraint language} to be an equality constraint language
where each relation is definable using equalities $(u = v)$ and the
\emph{positive} boolean connectives $\wedge$ and $\vee$.  As an example,
take the relation $S \subseteq D^4$ defined by
$$S(w, x, y, z) \equiv ((w = x) \wedge (x = y)) \vee (y = z).$$
It can be verified that a positive constraint language has all
unary operations as polymorphisms.

For a domain $D$ of any size, the CSP over a positive
constraint language is trivial; every instance is satisfiable
by the assignment sending all variables to the same value.
In contrast, the QCSP is more interesting.  We have shown that
there exist positive constraint languages $\Gamma$ over an infinite domain
such that $\qcsp(\Gamma)$ is NP-hard~\cite{bc-equality}; 
in fact, the constraint language
containing the single relation $S$ just given 
is an example of such a language.
This negative result is complemented by the following complexity upper bound.

\begin{theorem} \cite{bc06}
\label{thm:positive-in-np}
Let $\Gamma$ be a positive constraint language over an infinite domain,
and let 
$$\Phi = Q_1 v_1 \ldots Q_n v_n \phi$$
be an instance of $\qcsp(\Gamma)$.  The formula $\Phi$ is true if and only if
the formula
$$\Phi' = \exists v_1 \ldots \exists v_n (\phi \wedge \bigwedge_{i < j, Q_j = \forall}  v_i \neq v_j)$$
is true.
Hence, $\qcsp(\Gamma)$ reduces to $\csp(\Gamma \cup \{ \neq \})$, and is in NP.
\end{theorem}

I would like to highlight a
facet of our proof of Theorem~\ref{thm:positive-in-np}:
this proof \emph{only} uses the fact that the quantifier-free part $\phi$ of
a $\qcsp(\Gamma)$ instance $\Phi$ has all unary operations as polymorphisms.
We can thus use the same proof to establish an NP upper bound on
the more general class of quantified formulas where the quantifier-free
part is an \emph{arbitrary} positive formula!

\begin{theorem} (follows from~\cite{bc06})
\label{thm:in-np}
Let 
$$\Phi = Q_1 v_1 \ldots Q_n v_n \phi$$
be a formula where $\phi$ is composed from equalities $(u = v)$
and the connectives $\wedge$, $\vee$, and the variables are
interpreted over an infinite domain.
The formula $\Phi$ is true if and only if
the formula $\Phi'$ defined in the statement of 
Theorem~\ref{thm:positive-in-np} is true.
Hence, deciding formulas $\Phi$ of the described form is in NP.
\end{theorem}

It should be pointed out that the NP upper bound established by 
Theorems~\ref{thm:positive-in-np} and~\ref{thm:in-np} can be derived
from a result of Kozen~\cite{kozen-positive}.
However, I believe that the 
direct reduction to an existentially quantified formula
given here constitutes a
particularly transparent explanation for the inclusion in NP.
Again, as with the results discussed in the first section of this article,
these results 
are not tied to the particular syntactic form of
the quantifier-free part $\phi$, but only use the fact that $\phi$
possesses certain polymorphisms.
Indeed, our proof demonstrates that Theorem~\ref{thm:in-np} holds
even if the quantifier-free part $\phi$ is given succinctly, say, as
a circuit; this extension cannot, to the best of my knowledge,
be readily derived from the proof of Kozen~\cite{kozen-positive}.

What I personally find most interesting here is that we are able to obtain
a positive complexity result (Theorem~\ref{thm:in-np}) 
using the notion of polymorphism, 
for a class of formulas that falls \emph{beyond} the CSP framework,
that is, does not require a conjunction of constraints.
This is a twist of high intrigue: 
the concept of polymorphism has
recently come into close focus for its 
relevance to the \emph{specific} class of CSP formulas,
but has now been brought to 
shed light on a more \emph{general} class of logical formulas
(namely, the positive formulas of Theorem~\ref{thm:in-np}).
I would like to suggest the search for further applications of
polymorphisms to general logical formulas.

\paragraph{\bf Further reading.}  
For the reader interested in further studying the topics of this article,
I offer some pointers.  
I focus on the topics directly addressed by this article,
such as algebraic methods for studying the problems $\csp(\Gamma)$.
However,
even regarding these topics,
the selection and discussion of
references here is not at all meant to be comprehensive,
 but rather is a 
sampling of the literature that
reflects my personal interests and biases.

As we have discussed, Schaefer~\cite{schaefer78} was the first to
systematically consider the family of problems $\csp(\Gamma)$;
he studied constraint languages $\Gamma$ over a two-element domain.
While there does not appear to have been work on these problems
in the 80's, the early 90's saw papers
 of Hell and Nesetril~\cite{hn90} 
and Feder and Vardi~\cite{fv93,fv98} on these problems.
Feder and Vardi~\cite{fv93,fv98} conjectured that every problem of the form
$\csp(\Gamma)$ for $\Gamma$ over a finite domain is either in P or NP-complete;
this has become known as the Feder-Vardi dichotomy conjecture.
The algebraic, polymorphism-based approach 
to studying the problems $\csp(\Gamma)$ was
introduced in the papers of
Jeavons, Cohen, and Gyssens~\cite{jcg97} and Jeavons~\cite{jeavons98};
the given references are for journal papers
which appeared in the late 90's.
Other foundational work on this algebraic approach appears in
the journal paper by Bulatov, Jeavons, and Krokhin~\cite{bjk05}.

Two major complexity classification results 
on $\csp(\Gamma)$
were achieved by Bulatov: the classification of all
constraint languages $\Gamma$ over a 
three-element domain~\cite{bulatov06-dichotomy},
and the classification of all \emph{conservative} constraint languages
$\Gamma$, defined to be constraint languages containing 
all unary relations~\cite{bulatov03}.
There is by now a rich literature on $\csp(\Gamma)$ tractability
and complexity, including the 
papers~\cite{jcc98,dp99,bj03,bulatov04-graph,ktt04,bulatov05-revisited,dalmau05-gmm,bd-maltsev,kv06-cd,lz06-equations,lz-boundedwidth}.

Placing $\csp(\Gamma)$ problems in complexity classes ``below'' P has
also been studied.
Allender et al.~\cite{abisv05} gave a refinement of Schaefer's theorem
showing each problem $\csp(\Gamma)$, where
$\Gamma$ is a constraint language over a two-element domain,
to be complete for a known complexity class under AC$^0$
reductions.
See~\cite{dalmau05-linear,atserias05,llt06}
for examples of other papers along these lines.

A framework for studying the CSP over
\emph{infinite} domains was proposed by Bodirsky~\cite{bodirsky-thesis}; 
see~\cite{bodirsky05-cores,bd06-datalog,bk06}
 for subsequent work.
The quantified CSP has been studied algebraically in papers 
including~\cite{bbkj03,bbrcsv05,bc06,chen06-collapsibility,chen-existentially}.
For references on relevant algebra, we mention~\cite{szendrei86-clones,mmt87-book,bs-universalalgebra}.
Other surveys/overviews on these and related topics 
include~\cite{cks01,kbj03-ismvl,kbj03-sms,bcrv03-post,bcrv04-csp,cj06-chapter}.

\paragraph{\bf Acknowledgements.}
I extend special thanks to 
Manuel Bodirsky and
V\'{\i}ctor Dalmau for many enjoyable conversations on the topics
of this article.  I am also indebted to
Andrei Bulatov, Neil Immerman, Benoit Larose, Riccardo Pucella, Pascal Tesson,
and Matt Valeriote
for their numerous helpful comments and encouragement.

\end{document}